\pdfoutput=1
\documentclass[aps,prd,twocolumn,nofootinbib,longbibliography,amsfonts,amssymb,amsmath,superscriptaddress]{revtex4-2}
\usepackage[dvipsnames,table]{xcolor}
\usepackage{microtype}
\usepackage{IEEEtrantools}
\usepackage{mathrsfs}
\usepackage{amsthm}
\usepackage{enumitem}
\usepackage{varwidth}
\usepackage[normalem]{ulem}
\usepackage{graphicx}
\usepackage{stackengine}
\usepackage{cprotect}
\usepackage[caption=false]{subfig}
\usepackage{multirow}
\usepackage{hyperref}
\usepackage[USenglish]{babel}
\usepackage{microtype}
\usepackage{dashrule}

\newlength\replength
\newcommand\repfrac{.33}

\newcommand\rulewidth{.6pt}
\newcommand\tdashfill[1][\repfrac]{\cleaders\hbox to \replength{%
  \smash{\rule[\arraystretch\ht\strutbox]{\repfrac\replength}{\rulewidth}}}\hfill}

\usepackage{tikz}
\usetikzlibrary{shapes.geometric, arrows.meta, backgrounds,
  fit, graphs, quotes}
\makeatletter
\def\adl@drawiv#1#2#3{%
        \hskip.5\tabcolsep
        \xleaders#3{#2.5\@tempdimb #1{1}#2.5\@tempdimb}%
                #2\z@ plus1fil minus1fil\relax
        \hskip.5\tabcolsep}
\newcommand{\cdashlinelr}[1]{%
  \noalign{\vskip\aboverulesep
           \global\let\@dashdrawstore\adl@draw
           \global\let\adl@draw\adl@drawiv}
  \cdashline{#1}
  \noalign{\global\let\adl@draw\@dashdrawstore
           \vskip\belowrulesep}}
\makeatother

\tikzset{
  node distance=2cm,
  io/.style={trapezium, align=center, rounded corners, trapezium left angle=70,trapezium right angle=-70,minimum height=0.5cm, text centered, draw=black, fill=YellowOrange!20 },
  context/.style={trapezium, align=center, rounded corners, trapezium left angle=70,trapezium right angle=-70,minimum height=0.5cm, text centered, draw=black, fill=Cerulean!20 },
  process/.style={rectangle, align=center, minimum width=.75cm, minimum height=.75cm, text centered, draw=YellowOrange, line width=1.5pt},
  point/.style={circle,inner sep=0pt,minimum size=1pt, fill=black},
  op/.style={circle, minimum size=2pt, inner sep=0pt, text centered, draw=black},
  >={stealth},
  every new ->/.style={thick},
  flow/.style={thick, YellowOrange},
  line/.style={draw,thick,-stealth}
}

\usepackage[USenglish]{babel}
\usepackage{microtype}
\usepackage{amsmath}

\DeclareMathOperator*{\argmin}{arg\,min}
\DeclareMathOperator{\supp}{supp}

\begin{document}

\title{Adapting to noise distribution shifts in flow-based gravitational-wave inference}

\author{Jonas~Wildberger}
\email{wildberger.jonas@tuebingen.mpg.de}
\affiliation{Max Planck Institute for Intelligent Systems,  Max-Planck-Ring 4, 72076 T\"ubingen, Germany}
\author{Maximilian Dax}
\email{maximilian.dax@tuebingen.mpg.de}
\affiliation{Max Planck Institute for Intelligent Systems, Max-Planck-Ring 4, 72076 T\"ubingen, Germany}
\author{Stephen R. Green}
\email{stephen.green2@nottingham.ac.uk}
\affiliation{School of Mathematical Sciences, University of Nottingham\\ University Park, Nottingham NG7 2RD, United Kingdom}
\affiliation{Max Planck Institute for Gravitational Physics (Albert Einstein Institute), Am M\"uhlenberg 1, 14476 Potsdam, Germany}
\author{Jonathan Gair}
\affiliation{Max Planck Institute for Gravitational Physics (Albert Einstein Institute), Am M\"uhlenberg 1, 14476 Potsdam, Germany}
\author{Michael P\"urrer}
\affiliation{Max Planck Institute for Gravitational Physics (Albert Einstein Institute), Am M\"uhlenberg 1, 14476 Potsdam, Germany}
\affiliation{Department of Physics, East Hall, University of Rhode Island, Kingston, RI 02881, USA}
\affiliation{URI Research Computing, Tyler Hall, University of Rhode Island, Kingston, RI 02881, USA}
\author{Jakob H.~Macke}
\affiliation{Max Planck Institute for Intelligent Systems,  Max-Planck-Ring 4, 72076 T\"ubingen, Germany}
\affiliation{Machine Learning in Science, University of T\"ubingen, 72076 T\"ubingen, Germany}
\author{Alessandra Buonanno}
\affiliation{Max Planck Institute for Gravitational Physics (Albert Einstein Institute), Am M\"uhlenberg 1, 14476 Potsdam, Germany}
\affiliation{Department of Physics, University of Maryland, College Park, MD 20742, USA}
\author{Bernhard Schölkopf}
\affiliation{Max Planck Institute for Intelligent Systems,  Max-Planck-Ring 4, 72076  T\"ubingen, Germany}

\begin{abstract}

Deep learning techniques for gravitational-wave parameter estimation have emerged as a fast alternative to standard samplers---producing results of comparable accuracy. These approaches (e.g., \textsc{Dingo}) enable amortized inference by training a normalizing flow to represent the Bayesian posterior conditional on observed data. By conditioning also on the noise power spectral density (PSD) they can even account for changing detector characteristics. However, training such networks requires knowing in advance the distribution of PSDs expected to be observed, and therefore can only take place once all data to be analyzed have been gathered. Here, we develop a probabilistic model to forecast future PSDs, greatly increasing the temporal scope of \textsc{Dingo} networks. Using PSDs from the second LIGO-Virgo observing run (O2)---plus just a \emph{single} PSD from the beginning of the third (O3)---we show that we can train a \textsc{Dingo} network to perform accurate inference \emph{throughout} O3 (on 37 real events). We therefore expect this approach to be a key component to enable the use of deep learning techniques for low-latency analyses of gravitational waves.

\end{abstract}

\maketitle

\section{Introduction}
Detector noise plays a crucial role in interpreting observations of
gravitational waves (GWs). In its simplest form, noise is assumed to be
additive, and stationary and Gaussian. This means that it is
characterized in frequency domain by its power spectral density (PSD)
$S_\text{n}(f)$. The GW likelihood for parameters $\theta$ is then the
probability that after a proposed signal $h(\theta)$ is subtracted
from data $d$, the residual is noise satisfying these assumptions,
i.e.,
\begin{equation}
  p(d | \theta, S_\text{n}) \propto \exp\left( - \frac{1}{2} (d - h(\theta) | d - h(\theta))_{S_\text{n}} \right),
\end{equation}
where $(\cdot | \cdot)$ is the noise-weighted inner-product,
\begin{equation}
  (a | b)_{S_\text{n}} = 4 \sum_I \Re \int_{f_{\text{min}}}^{f_{\text{max}}} \frac{\hat a_I^\ast (f) \hat b_I(f)}{S_{\text{n},I}(f)} df,
\end{equation}
and the sum runs over interferometers $I$.

Although the LIGO~\cite{TheLIGOScientific:2014jea},
Virgo~\cite{TheVirgo:2014hva} and KAGRA~\cite{Somiya:2011np,Aso:2013eba,KAGRA:2020tym} detectors are mostly stable during an
observing run, the noise spectrum does vary over time and across
detectors. Moreover, between observing runs, detectors are upgraded,
resulting in reduced noise levels and increased sensitivity. The
particular PSDs at the time of an event must therefore be estimated
and taken into account when performing inference. Estimation of a PSD
is typically carried out using either data adjacent to an event
(off-source, Welch method~\cite{welch1967use}) or by jointly modeling the signal and noise
from the on-source data (e.g.,
\textsc{BayesWave}~\cite{Cornish:2014kda,Cornish:2020dwh}). The PSD is
then inserted into the likelihood and samples are drawn from the posterior
$p(\theta|d,S_\text{n})$ using Markov chain Monte
Carlo~\cite{Veitch:2014wba} or nested sampling
methods~\cite{Ashton:2018jfp,Romero-Shaw:2020owr,Speagle_2020}.

An emerging alternative to classical likelihood-based inference is
simulation-based inference using probabilistic deep
learning~\cite{Gabbard:2019rde,Chua:2019wwt,Chatterjee:2019gqr,Green:2020hst,Green:2020dnx,Delaunoy:2020zcu,Dax:2021tsq,Dax:2021myb,Krastev:2020skk,Shen:2019vep}. Technologies
such as normalizing flows enable neural networks to describe complex
conditional probability distributions. A conditional density estimator
$q(\theta|d)$ can be trained using simulated data to approximate
$p(\theta|d)$ such that once a detection is made, samples can be drawn
in seconds. To incorporate varying noise PSDs into these methods, the
estimate of $S_\text{n}$ is provided as additional context to the
network, i.e.\ $q(\theta|d, S_\text{n})$. During training, random PSDs
are drawn from an empirical distribution $p(S_\text{n})$ estimated
from signal-free data during an observing run. Simulated data are then
constructed based on these PSDs, and the PSD is provided as
context. At inference time, the network is effectively ``tuned'' to
the interferometers at the time of detection. This approach (called
\textsc{Dingo}~\cite{Dax:2021tsq}) fully amortizes training costs
across all detections within an observing run, and has been shown to
produce results nearly indistinguishable from classical samplers.

\begin{figure}%
\centering
\includegraphics[width=.45\textwidth]{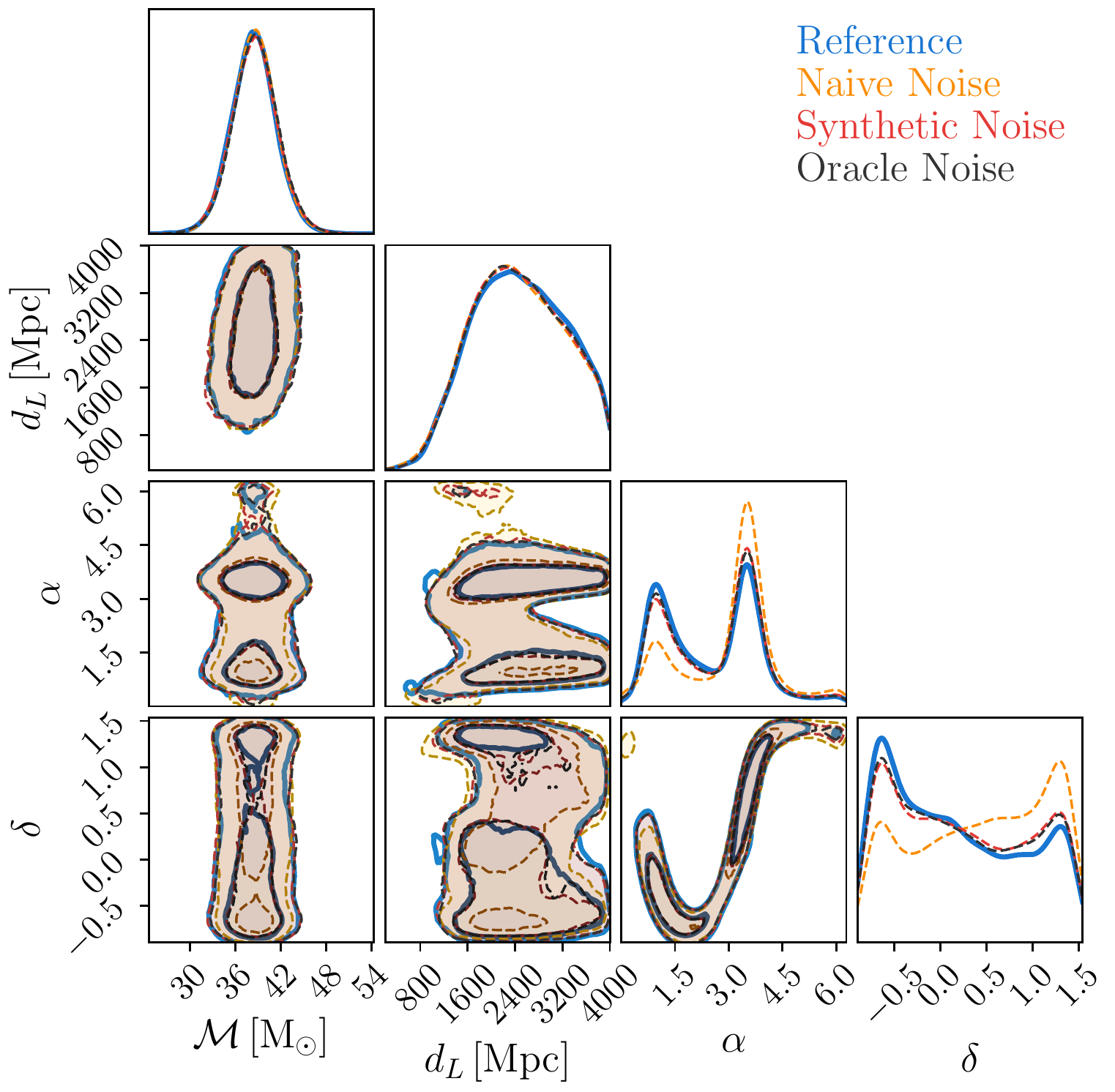}
\caption{
Posterior for GW200208\_130117. Results from \textsc{Dingo} models trained only with empirically estimated detector-noise PSDs from the beginning of an observing run (orange) may deviate visibly from the reference (blue, obtained using importance sampling~\cite{Dax:2022pxd}). \textsc{Dingo} models trained with our proposed synthetic PSD model (red) achieve much more accurate results, on par with using PSDs estimated throughout the entire observing run (black). Our PSD model therefore enables \textsc{Dingo} to adapt to unseen drifts of the PSDs without retraining. 
}
\label{fig:GW200208_130117}
\end{figure}

The approach described requires access to the empirical distribution
$p(S_\text{n})$ of PSDs during an observing run. This makes it
unsuitable for \emph{online} parameter estimation (e.g., to provide
alerts to electromagnetic telescopes) since the distribution $p(S_\text{n})$
covering future observations is unavailable at the time of network
training. Thus, a network trained with an empiric PSD distribution can
only be used for a limited time---once the PSDs change too much, the
measured data become out-of-distribution (OOD). Such a disagreement
between training and inference distributions can lead to inaccurate
results (Fig.~\ref{fig:GW200208_130117}). Here, we address this
problem and provide a solution for training \textsc{Dingo} models
robustly to better adapt to shifting distributions.

\begin{figure}
  \centering
  \includegraphics[width=0.48\textwidth]{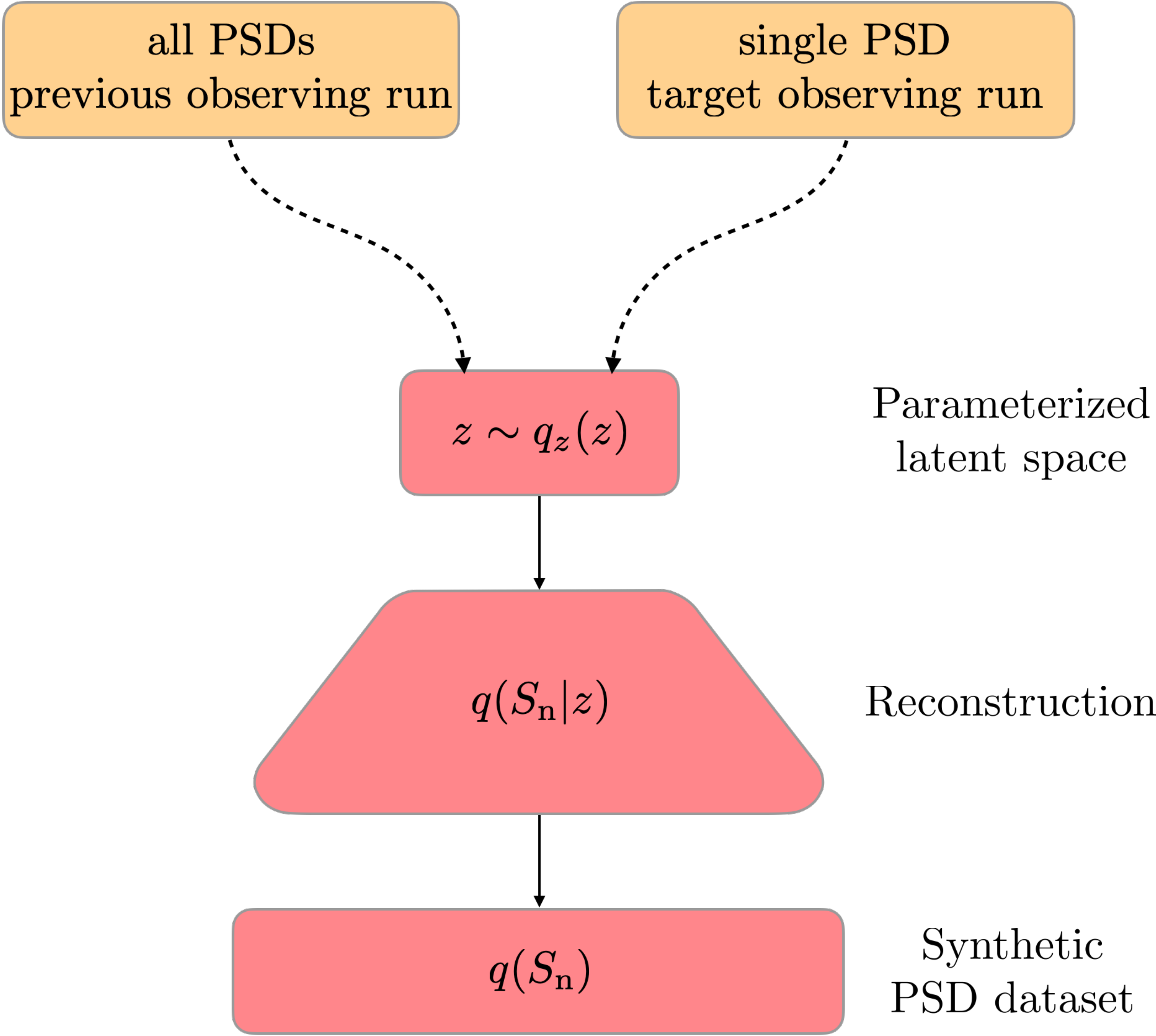}
  \caption{Overview of the PSD generation pipeline. The PSDs from the previous observing run and a single PSD from the target observing run are used for fitting the latent distribution $q_z(z)$, as indicated by the dashed arrows. We can then generate synthetic PSDs
  by first sampling latent variables in $z\sim q_z(z)$. These are then reconstructed back to frequency space via $q(S_\text{n} | z)$, obtaining a synthetic PSD distribution $q(S_\text{n})$. }
  \label{fig:overview_method}
\end{figure}

Our approach is based on the key property that \textsc{Dingo} models
$q(\theta | d, S_\text{n})$ estimate a distribution \emph{conditional
on} (as opposed to \emph{marginalized over}) the PSD $S_\text{n}$. We
are therefore not restricted to using (an approximation to) the real
distribution $p(S_\text{n})$ of PSDs; instead, we can use a synthetic
distribution $q(S_\text{n})$ whose support \emph{contains} that of $p(S_\text{n})$. In other words, if $\supp p(S_\text{n})\subseteq \supp q(S_\text{n})$, then the
\textsc{Dingo} model trained with $q(S_\text{n})$ can be used for the entire observing run. 

In this work, we develop a parameterized latent variable model for $q(S_\text{n})$, which we fit to PSDs from a previous observing run. It is then straightforward to modify this distribution of PSDs via operations on the latent space. In particular, we use a one-shot observation from an upgraded detector to shift the latent space distribution coarsely to the expected noise level. Further, we broaden the spread of PSDs by blurring the latent space distribution. Our model $q(S_\text{n})$ thereby represents a broad distribution over PSDs, which is capable of capturing variations throughout an observing run. %

We evaluate our approach on the third LIGO-Virgo-KAGRA (LVK) observing run (O3) by analyzing 390 simulated and 37 real GW events~\citep{Abbott:2019ebz}. We train \textsc{Dingo} with our PSD model and consistently find similar performance to \textsc{Dingo} trained with real O3 PSDs. Since our PSD model only uses O2 data and a single PSD from the beginning of O3, this demonstrates that our approach is indeed capable of preparing \textsc{Dingo} for unseen PSD changes.

\section{Methods}\label{sec:methods}
In probabilistic modeling, a latent variable model~\cite{bartholomew2011latent,Kingma2014,rezende2014stochastic} enables efficient sampling of new data given an empirical distribution. We define a latent variable model for the PSDs, as
\begin{equation}
	q(S_\text{n}) = \int q(S_\text{n}|z) q_z(z) dz.
\end{equation}
We use a fixed parameterization for $q(S_\text{n}|z)$ to integrate knowledge 
about the data generating process into the model. PSD data can then be fitted via the distribution $q_z(z)$ over the latent variables. This provides an interpretable framework, enabling systematic interventions on the PSD distribution. Synthetic PSDs can then be generated by sampling $z\sim q_z(z)$ and reconstructing $z$ back to frequency space via $q(S_\text{n} | z)$. This pipeline is visualized in Fig.~\ref{fig:overview_method}; next we explain the individual components in detail.

\begin{figure}%
\centering
\includegraphics[width=.48\textwidth]{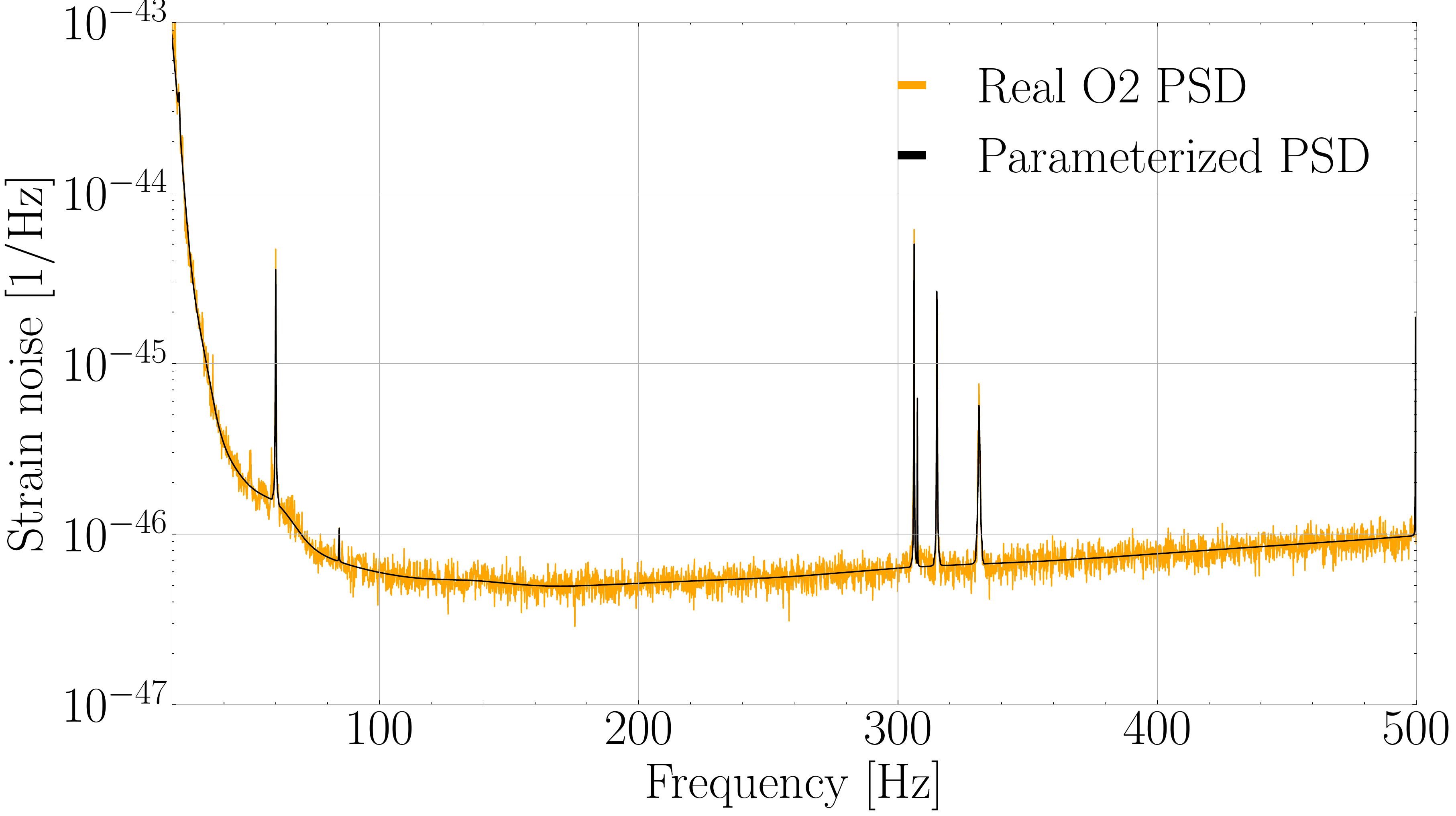}\vspace{0.2cm}
\includegraphics[width=.48\textwidth]{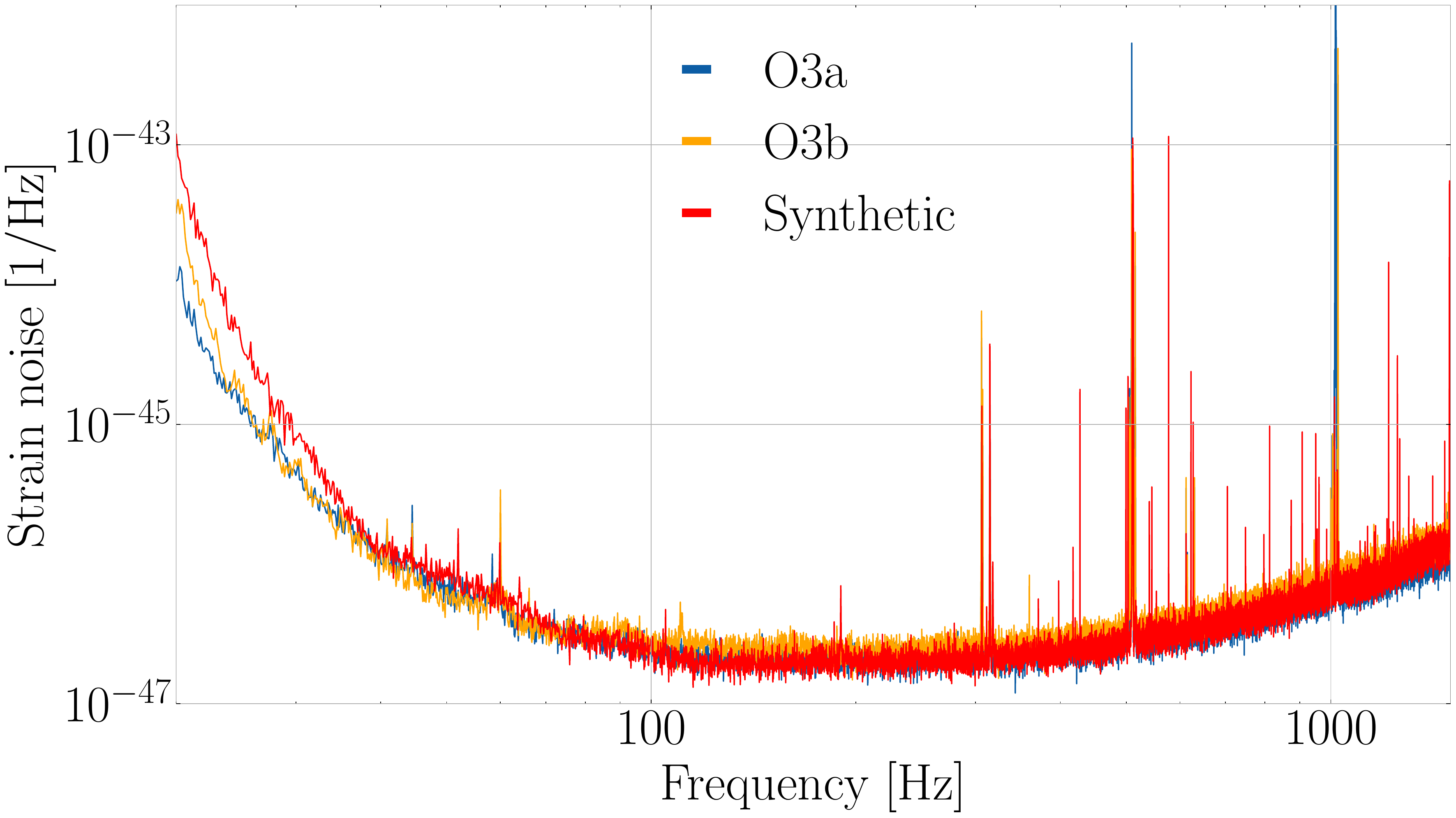}
\caption{
Upper panel: Comparison of a real O2 PSD and its representation under the latent variable model. Since $l=400$ is sufficiently large, we can accurately model spectral lines that are close together (at around 300Hz). Lower panel: Comparison of a synthetically sampled PSD and two real PSDs from O3a and O3b. By shifting the broad-band noise level (\ref{sec:shifting}), we match the overall scale of O3 PSDs. By broadening the distribution, we ensure that PSDs from both O3a and O3b are covered.
}
\label{fig:comparison_PSDs}
\end{figure}

\subsection{Parametrization of $q(S_\text{n}|z)$}

We aim to design a model $q(S_\text{n})$ with sufficient expressiveness to capture every realistic PSD. On the other hand, the latent space should be low dimensional and maximally disentangled, such that the distribution $q_z(z)$ partially factorizes. In particular, the latent features $z$ should model those PSD characteristics that are likely to change over time. This increases the data efficiency and robustness of our model under interventions. These requirements inform our design of $q(S_\text{n}|z)$. 

We leverage domain knowledge to parameterize the latent feature space explicitly. Each PSD is represented by a $n$-dimensional vector $S_\text{n} \in \mathbb{R}^n$ for uniformly distributed frequency bins $f_1, \ldots, f_n$ in the range $[f_\text{min}, f_\text{max}]$.  According to~\citep{Cornish:2014bl}, $S_\text{n}$ can be decomposed into two main components, the smooth broad-band noise $b \in \mathbb{R}^n$ and a sum of high-power spectral lines $\sum_i s_i\in \mathbb{R}^n$. Improvements to the detectors are intended to reduce the impact of the broad-band noise, making it likely to change over time. The position and shape of the spectral lines may vary on much shorter time-scales. We thus model these components independently. We further assume independent additive Gaussian noise with constant variance $\sigma^2$ in each frequency bin.

With these assumptions, our model reads
\begin{equation}\label{eq:PSD_model}
    q(S_\text{n} | z) = \mathcal N\left(b + \sum_{i=1}^l s_i, \sigma^2 I_n\right).
\end{equation}

In latent space, the broad-band noise is represented by $k$ values $y_1, \ldots, y_k\in \mathbb{R}_+$ on fixed log-spaced frequency nodes $x_1, \ldots, x_k$. The logarithmic distribution accounts for larger fluctuations of the broad-band noise in lower frequency regions. $b$ is then obtained from its latent representation via cubic spline interpolation between the nodes.

Each spectral line $s_i$ is represented by parameters $f_{0i}, A_i, Q_i$, denoting the center, height, and width of a truncated Cauchy distribution, respectively, i.e.\,
\begin{equation}
    s_i[f_{0i}, A_i, Q_i](f) = \frac{z(f, f_{0i}) A_i f_{0i}^4}{(f_{0i} f)^2 + (Q_i(f_{0i}^2 - f^2))^2},
\end{equation}
with 
\begin{equation}
    z(f, f_{0i}) = \begin{cases}
        1, & \text{if } |f - f_{0i}| \leq \delta, \\
        \exp(- \frac{|f - f_{0i}|}{\delta}), & \text{otherwise},
    \end{cases}
\end{equation}
similarly to~\cite{Cornish:2014bl}. To efficiently treat a varying number of spectral lines per PSD, we segment the frequency range into $l$ equally-wide intervals and model a single spectral line within each. Absence of a spectral line is modeled with $A_i\approx 0$. 
We do not find that restricting to one line per interval hinders \textsc{Dingo} performance in practice, provided the number of intervals is sufficiently large; see Fig.~\ref{fig:comparison_PSDs} (upper panel). In our experiments, we used $k=30$, and $l=400$ over a frequency range of $[f_\text{min}, f_\text{max}] = [20, 2048]~
\text{Hz}$. If desired, the model could be extended to use overlapping intervals. In this case, each spectral distribution would no longer be independent, but rather conditional on the parameters of the preceding frequency interval.

We found that $\sigma^2$ does not vary significantly between different PSDs. We therefore use a fixed value for $\sigma^2$ (estimated from the variance of the observed broad-band noise), and do not consider it as part of our latent space. Summarizing, our mapping from the latent space to frequency space reads
\begin{equation}
   z = \begin{pmatrix}
   y_1, \ldots, y_k, \\
   f_{01}, A_1, Q_1,\\
   \vdots  \\
   f_{0l}, A_l, Q_l
   \end{pmatrix} \mapsto S_\text{n}.
\end{equation}
The dimensionality of our latent space is $d=k + 3l$. %

\subsection{Fitting the latent distribution $q_z(z)$}

We fit $q_z(z)$ based on an empirical distribution of PSDs estimated from detector data. We estimate this collection of PSDs using Welch's method applied to signal-free data stretches during the previous observing run. We then project these onto the latent space using maximum likelihood estimation and obtain $q_z(z)$ as the product of Gaussian kernel density estimates (KDEs) of these projections. This gives a tractable approximation to the true latent distribution. Importantly, having an analytic KDE provides flexibility to widen $q_z(z)$ by increasing the KDE bandwidth.

We perform the maximum likelihood projection onto the latent space in two steps. We first estimate the spline parameters $y_1, \ldots, y_k$ for the broad-band noise as the sample mean of the PSD in the vicinity of the corresponding $x_i$. Since the broad-band noise should not model the spectral lines, we apply a filter before this step.\footnote{We compute a running median over the neighborhood sets of each $x_i$. Every data point that deviates by more than three standard deviations from the running median is marked as an outlier and ignored for the fit of the broad-band noise.} Once the broad-band noise is fitted, we subtract it from the PSD to fit the spectral lines. Specifically, we obtain $(f_{0i}, A_i, Q_i)$ by solving the least-squares problem
\begin{align}
\begin{split}
    (f_{0i}, A_i, Q_i) &= \argmin_{(\tilde f, \tilde A, \tilde Q)} \sigma^{-2}r r^\top \\
    r = r (f, A, Q) &= S_{\text{n}} - (b + s_i(f, A, Q)),
\end{split}
\end{align}
over each interval $i$.
Together, these two steps correspond to a maximum-likelihood estimate of $z$ for \eqref{eq:PSD_model} with fixed $\sigma^2$. 
On eight CPU cores, it takes less than a minute to obtain the latent maximum likelihood estimate $z_\text{MLE}$ for each PSD, and this procedure is straightforwardly parallelizable. Indeed, this fast calculation (compared to $\approx 1$~hour for BayesLine) is one reason for choosing this simpler approach when building our PSD model.

Having computed $z_\text{MLE}$ for each PSD from the empiric distribution, we next use KDEs to obtain $q_z(z)$. We use a separate KDE for each "independent" noise source, and then combine them multiplicatively. For the broad-band noise, we partition the frequency range into three intervals according to the most dominant noise source: $\mathcal F_1 = [f_{\text{min}}, 30~\text{Hz}]$ for seismic noise, $\mathcal F_2 = [30~\text{Hz}, 100~\text{Hz}]$ for thermal noise and $\mathcal F_3 = [100~\text{Hz}, f_{\text{max}}]$ for shot noise~\cite{Cornish:2014bl}. Since these noise sources should be largely independent of each other, we use individual KDEs for each subset $\{y_i \ | \ x_i \in \mathcal F_j\}$. These subsets do not overlap, so the reconstructed spline naturally agrees on the boundary between intervals $\mathcal F_i$.
Similarly, the spectral lines are modeled independently,
so we use a separate KDE for each set of parameters $(f_{0i}, A_i, Q_i)$. The latent distribution is then given as the product of the $3 + l$ individual KDEs.

The independence assumption made above, and the corresponding partial factorization of $q_z(z)$ into $3 + l$ independent distributions, serves two purposes. First, it provides broader coverage of the latent space compared to an unfactorized KDE. Second, it decorrelates factors of variation that are independent so that trained networks do not learn to expect spurious correlations that exist in the empirical PSDs. Both of these effects ensure that trained networks are generally more robust with respect to changes in the PSD so that they can adapt to unseen data.

\subsection{Interventions in latent space}\label{sec:shifting}

The latent space is much lower dimensional than the raw PSD space and has features that correspond to the main components comprising a PSD. By intervening in this space (i.e., changing the PSD distribution at the level of the latent space) we can therefore naturally adapt the PSD distribution to changing detector noise characteristics. We perform two such types of interventions: (1) we shift the broad-band noise to account for improved detector sensitivity, and (2) we broaden the distribution to account for uncertainty in the estimated broad-band shift and other unmodeled detector changes. 

For (1), we rescale the latent distribution over $y_1, \ldots, y_k$ according to an estimated PSD from early in the target run.
Given a single PSD $S_\text{n}^\text{target}$ from the target run, we shift the distribution over $y_i$ so that its mean corresponds to the target PSD, i.e.,
\begin{equation}
    y_i \mapsto y_i - \mathbb{E}[y_i] + y_i^\text{target}.
\end{equation}
Here, $y_i^\text{target}$ refers to the latent features corresponding to $S_n^\text{target}$. The resulting distribution corresponds to variations in all latent variables inferred from past data, but with broad-band noise level shifted to that of the target observing run. Note that this works even when $S_\text{n}^\text{target}$ is not close to the mean of the target run, as long as the difference is small compared to the variance of the estimated distribution. 
In Fig.~\ref{fig:comparison_PSDs}, we illustrate how this intervention on the latent space ensures that synthetically generated PSDs match the broad-band noise level of O3 PSDs (lower panel), despite the difference in scale between O2 and O3.

The broadening (2) is controlled by the KDE bandwidth parameter, and it improves the robustness of \textsc{Dingo} networks trained with the synthetic noise distribution. A larger bandwidth parameter can compensate for more significant shifts in the PSD distribution, although it may require higher learning capacity.

\section{Results}
\begin{figure*}
\centering
\includegraphics[width=\textwidth]{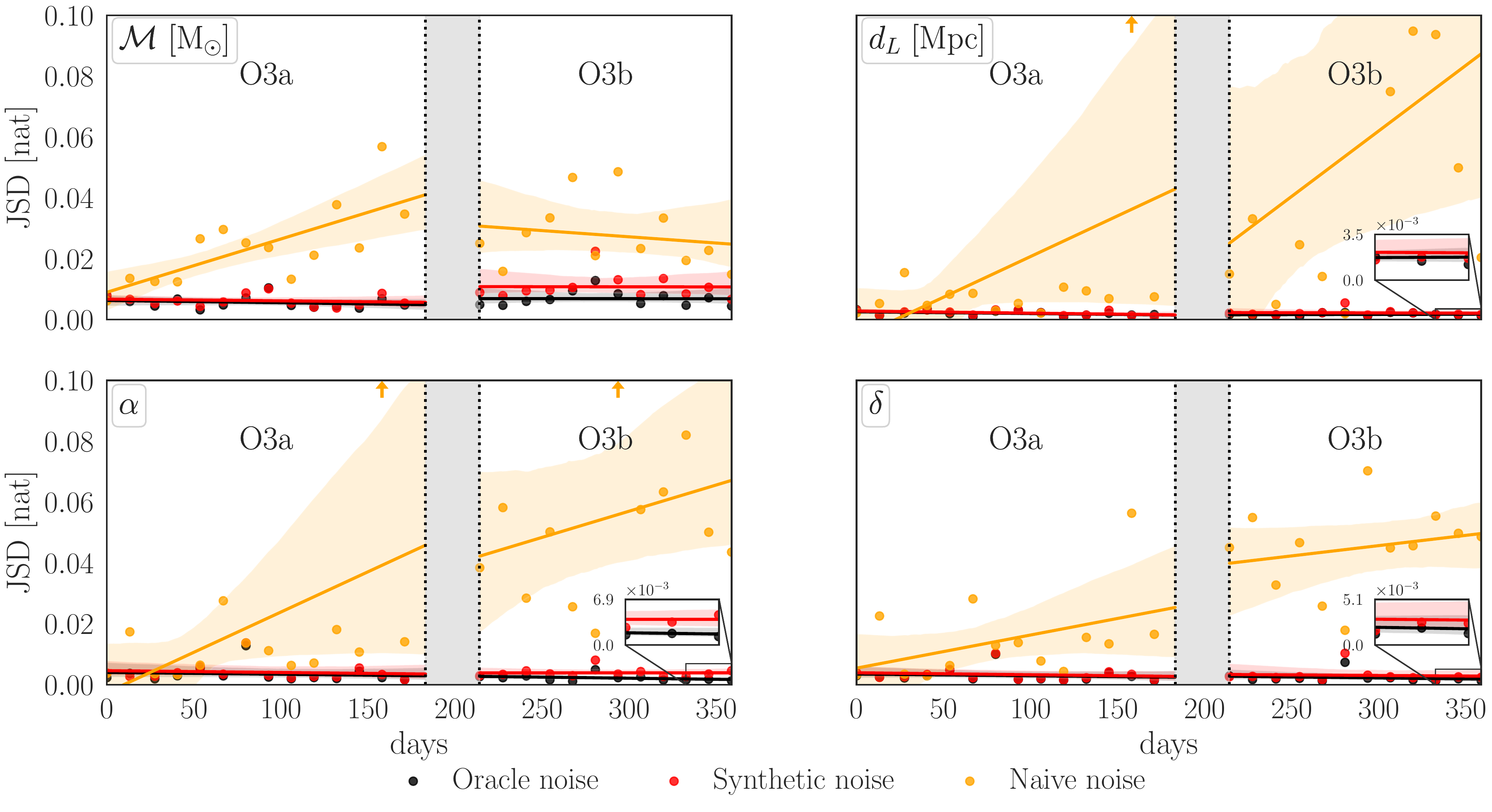}
\caption{
JSDs between \textsc{Dingo} and reference samples. Injections use noise PSDs estimated at various times throughout O3 (indicated on horizontal axis). Results are averaged over injections with five random sets of source parameters (fixed for all PSDs). Day 0 marks the beginning of O3a. The gap indicates the period when detectors were offline between O3a and O3b. 
}
\label{fig:injection_study}
\end{figure*}

\begin{figure*}

\centering
\includegraphics[width=\textwidth]{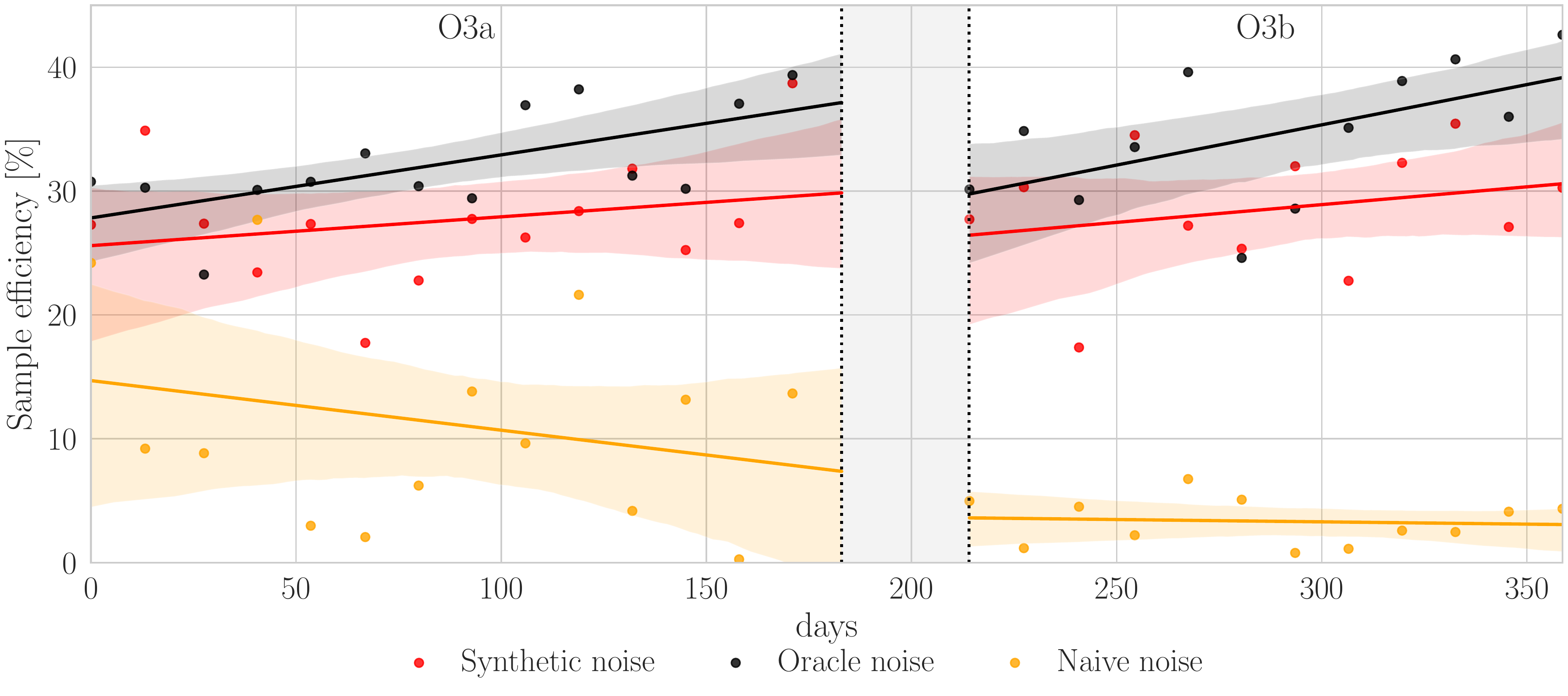}

\caption{
Importance sampling efficiency $\epsilon$ for \textsc{Dingo} models. Injections use noise PSDs estimated at various times throughout O3 (indicated on horizontal axis). Results are averaged over injections with five random sets of source parameters (fixed for all PSDs). Day 0 marks the beginning of O3a. The gap indicates the period when detectors were offline between O3a and O3b.
}
\label{fig:injection_study_ESS}

\end{figure*}

We prepare \textsc{Dingo} networks with the settings from \cite{Dax:2021tsq}. In particular, we use the waveform model IMRPhenomPv2 \cite{Khan:2015jqa, Hannam:2013oca, Bohe:2016gbl} with frequency-domain data in the range $[20,1024]$~Hz, $\Delta f = 0.125$~Hz, and a uniform distance prior of $[0.1,2]$~Gpc. 
We train three networks, each with a different noise PSD dataset: (1) the \emph{Oracle} dataset consists of PSDs estimated from real O3a and O3b detector noise ($\sim$5,000 PSDs per detector);
(2) the \emph{Synthetic} dataset is sampled with our proposed method, using only O2 data and a single PSD from the start of O3 (50,000 PSDs per detector); and (3) the \emph{Naive} baseline dataset consists of real PSDs from the first four days of O3a (100 PSDs per detector).
The \textit{Oracle} dataset encompasses the real PSDs that \textsc{Dingo} encounters at inference time, so this should be an upper bound for the performance of our method. On the other hand, the \textit{Naive} dataset is based only on data available at the beginning of O3, so to be useful our method should significantly outperform this baseline.

We evaluate the three \textsc{Dingo} networks on simulated and real strain data. To assess performance, we compare against reference posteriors. Since we analyze more than 400 events, generating these using stochastic samplers is not feasible. Instead, we generate reference posteriors by importance-weighting the \textsc{Dingo} results produced with the \textit{Oracle} PSD dataset, which provides verified inference results at low computational cost (\textsc{Dingo-IS})~\cite{Dax:2022pxd}.
To save computational time, we further use phase marginalization when calculating the likelihoods~\cite{veitch2013analytic,Veitch:2014wba,Thrane:2018qnx}. We note that with precessing waveforms this is only an approximation, but for IMRPhenomPv2 the error introduced is small. See \cite{Dax:2022pxd} for an alternative approach to generate the phase in the presence of higher modes.

\subsection{Simulated data}\label{sec:sim-data}

We first evaluate inference results when solely the PSD is varied in the strain data. To this end, we sample gravitational-wave parameters from the prior and inject the signal into identical noise realizations that are scaled with different PSDs. Specifically, we choose 26 real PSDs from O3, evenly spaced in time over the course of one year. The injection strains are then analyzed using each of the three \textsc{Dingo} networks.

Fig.~\ref{fig:injection_study} shows the mean Jensen-Shannon divergence (JSD)~\cite{Lin:1991zzm} between \textsc{Dingo} and \textsc{Dingo-IS} posteriors for chirp mass ($\mathcal{M}$), luminosity distance ($d_\text{L}$) and sky position ($\alpha,\delta$). These parameters have been found to be most significantly impacted by conditioning the network on an incorrect PSD at inference time~\cite{Dax:2021tsq}. (We find a similar trend for training with incomplete PSD information.) Results for all 15 parameters are provided in the appendix, showing a similar qualitative behavior to the parameters reported here.

At the beginning of the observing run, we observe similar accuracy for all three networks. This is not surprising, as the PSDs are still captured by the \textit{Naive} dataset at this point. However, as PSDs drift away from their initial distribution, the performance of the \textit{Naive} baseline decreases substantially, with JSDs increasing by one to two orders-of-magnitude. The effect is particularly striking at the transition from O3a to O3b. This demonstrates that the \textit{Naive} baseline indeed fails to generalize well to unseen PSDs. The \textit{Oracle} network on the other hand shows excellent performance throughout the entire duration of O3, with JSDs around $2\cdot 10^{-3}$~nat.\footnote{For comparison, a maximum JSD of $2\cdot 10^{-3}$~nat has been established as a bound for indistinguishability in~\cite{2020MNRAS.499.3295R}.} This is also expected, since the \textit{Oracle} network has access to PSDs from the entire duration of O3 during training.

Finally, with the \textit{Synthetic} PSD dataset, \textsc{Dingo} accuracy approaches that of the \textit{Oracle} network---even for PSDs recorded almost a year after the PSD used to recalibrate the \textit{Synthetic} dataset (zoomed-in parts in Fig.~\ref{fig:injection_study}). It far outperforms the \textit{Naive} baseline, demonstrating that with our proposed method, \textsc{Dingo} can indeed be trained to generalize to unseen PSDs, with at most a small decrease in accuracy.

In the same experiment, we study the sample efficiency $\epsilon$ when using the inferred \textsc{Dingo} posterior as a proposal distribution for importance sampling~\cite{Dax:2022pxd} (Fig.~\ref{fig:injection_study_ESS}). 
Small values of $\epsilon$ have been found to flag failure cases of the inference network, e.g.,~caused by out-of-distribution data, establishing it as another quality measure of the \textsc{Dingo} posteriors.
For the \textit{Naive} baseline, $\epsilon$ decreases with time and plummets after the transition from O3a to O3b. In contrast, the \textit{Oracle} and \textit{Synthetic} networks provide a consistently high sample efficiency of $\epsilon\approx 30\%$. As a performance metric, $\epsilon$ is sensitive to deviations in full parameter space, including high-dimensional correlations. Fig.~\ref{fig:injection_study_ESS} thus confirms the trend observed for the JSDs, and demonstrates that the \textit{Synthetic} PSD dataset can be used to ensure a high efficiency of \textsc{Dingo-IS} throughout an observing run.

\begin{table*}[]
  \centering
  \begin{tabular}{lllr}\hline\hline
        & \multicolumn{3}{c}{\bf{Noise datasets}} \\
         Event & Oracle & Synthetic & Naive\\\hline\hline
        
GW190408\_181802 & \cellcolor[HTML]{519F78} $2.1$ & \cellcolor[HTML]{FEFFD1} $4.0$ & \cellcolor[HTML]{C4E699} $3.1$\\ 
GW190413\_052954 & \cellcolor[HTML]{4D9573} $0.2$ & \cellcolor[HTML]{4D9573} $0.5$ & \cellcolor[HTML]{4D9573} $0.4$\\ 
GW190413\_134308 & \cellcolor[HTML]{4D9573} $0.9$ & \cellcolor[HTML]{4D9573} $0.6$ & \cellcolor[HTML]{4D9573} $2.0$\\ 
GW190421\_213856 & \cellcolor[HTML]{4D9573} $0.3$ & \cellcolor[HTML]{4D9573} $0.4$ & \cellcolor[HTML]{4D9573} $0.5$\\ 
GW190503\_185404 & \cellcolor[HTML]{4D9573} $1.1$ & \cellcolor[HTML]{97D292} $2.7$ & \cellcolor[HTML]{4D9573} $1.6$\\ 
GW190513\_205428 & \cellcolor[HTML]{509C76} $2.1$ & \cellcolor[HTML]{F0F9BE} $3.7$ & \cellcolor[HTML]{AEDC95} $2.9$\\ 
GW190514\_065416 & \cellcolor[HTML]{4D9573} $0.5$ & \cellcolor[HTML]{4D9573} $0.5$ & \cellcolor[HTML]{4D9573} $0.7$\\ 
GW190517\_055101 & \cellcolor[HTML]{C04D67} $21.3$ & \cellcolor[HTML]{C04D67} $25.4$ & \cellcolor[HTML]{C04D67} $14.2$\\ 
GW190519\_153544 & \cellcolor[HTML]{4D9573} $1.2$ & \cellcolor[HTML]{4D9573} $1.4$ & \cellcolor[HTML]{4D9573} $1.2$\\ 
GW190521\_074359 & \cellcolor[HTML]{4D9573} $1.4$ & \cellcolor[HTML]{6CBE88} $2.4$ & \cellcolor[HTML]{99D392} $2.7$\\ 
GW190527\_092055 & \cellcolor[HTML]{FCBB8B} $5.5$ & \cellcolor[HTML]{53A179} $2.1$ & \cellcolor[HTML]{4D9573} $1.4$\\ 
GW190602\_175927 & \cellcolor[HTML]{4D9573} $1.9$ & \cellcolor[HTML]{4D9573} $1.9$ & \cellcolor[HTML]{D5EDA4} $3.3$\\ 
GW190701\_203306 & \cellcolor[HTML]{4D9573} $1.2$ & \cellcolor[HTML]{4D9573} $1.4$ & \cellcolor[HTML]{4D9573} $1.4$\\ 
GW190719\_215514 & \cellcolor[HTML]{4D9573} $0.7$ & \cellcolor[HTML]{4D9573} $0.3$ & \cellcolor[HTML]{4D9573} $0.4$\\ 
GW190727\_060333 & \cellcolor[HTML]{4D9573} $0.3$ & \cellcolor[HTML]{4D9573} $0.6$ & \cellcolor[HTML]{4D9573} $0.6$\\ 
GW190731\_140936 & \cellcolor[HTML]{4D9573} $0.3$ & \cellcolor[HTML]{4D9573} $0.4$ & \cellcolor[HTML]{4D9573} $1.1$\\ 
GW190803\_022701 & \cellcolor[HTML]{4D9573} $0.6$ & \cellcolor[HTML]{4D9573} $0.3$ & \cellcolor[HTML]{4D9573} $1.4$\\ 
GW190805\_211137 & \cellcolor[HTML]{4D9573} $0.3$ & \cellcolor[HTML]{4D9573} $0.4$ & \cellcolor[HTML]{4D9573} $1.1$\\ 
GW190828\_063405 & \cellcolor[HTML]{4D9573} $0.7$ & \cellcolor[HTML]{55A57B} $2.1$ & \cellcolor[HTML]{89CC8F} $2.6$\\ \hline
  \end{tabular}\hspace{0.5cm}
 \begin{tabular}{lllr}\hline\hline
        & \multicolumn{3}{c}{\bf{Noise datasets}} \\
         Event & Oracle & Synthetic & Naive\\\hline\hline
GW190909\_114149 & \cellcolor[HTML]{4D9573} $0.4$ & \cellcolor[HTML]{4D9573} $0.5$ & \cellcolor[HTML]{4D9573} $0.9$\\ 
GW190915\_235702 & \cellcolor[HTML]{4D9573} $0.8$ & \cellcolor[HTML]{4D9573} $0.7$ & \cellcolor[HTML]{4D9573} $1.4$\\ 
GW190926\_050336 & \cellcolor[HTML]{4D9573} $1.7$ & \cellcolor[HTML]{4D9573} $1.8$ & \cellcolor[HTML]{4D9573} $1.8$\\ \hline
\multicolumn{1}{c}{\bf{O3b} $\downarrow$} & & & 
\\\hline
GW191127\_050227 & \cellcolor[HTML]{E3F3AD} $3.5$ & \cellcolor[HTML]{4D9573} $1.7$ & \cellcolor[HTML]{D1EBA1} $3.2$\\ 
GW191204\_110529 & \cellcolor[HTML]{C8E89C} $3.1$ & \cellcolor[HTML]{FFF0B9} $4.4$ & \cellcolor[HTML]{F7997B} $6.1$\\ 
GW191215\_223052 & \cellcolor[HTML]{4D9573} $0.8$ & \cellcolor[HTML]{4D9573} $1.1$ & \cellcolor[HTML]{A2D793} $2.8$\\ 
GW191222\_033537 & \cellcolor[HTML]{4D9573} $0.4$ & \cellcolor[HTML]{4D9573} $0.5$ & \cellcolor[HTML]{4D9573} $1.4$\\ 
GW191230\_180458 & \cellcolor[HTML]{4D9573} $0.3$ & \cellcolor[HTML]{4D9573} $0.4$ & \cellcolor[HTML]{4D9573} $0.7$\\ 
GW200128\_022011 & \cellcolor[HTML]{4D9573} $0.6$ & \cellcolor[HTML]{4D9573} $1.0$ & \cellcolor[HTML]{ACDC94} $2.9$\\ 
GW200129\_065458 & \cellcolor[HTML]{92D091} $2.6$ & \cellcolor[HTML]{4D9573} $1.6$ & \cellcolor[HTML]{FDC48F} $5.3$\\ 
GW200208\_130117 & \cellcolor[HTML]{4D9573} $0.5$ & \cellcolor[HTML]{4D9573} $0.6$ & \cellcolor[HTML]{FCB98A} $5.5$\\ 
GW200208\_222617 & \cellcolor[HTML]{5DB483} $2.3$ & \cellcolor[HTML]{5BB181} $2.3$ & \cellcolor[HTML]{C7E79B} $3.1$\\ 
GW200209\_085452 & \cellcolor[HTML]{4D9573} $1.3$ & \cellcolor[HTML]{4D9573} $1.8$ & \cellcolor[HTML]{5BB181} $2.2$\\ 
GW200216\_220804 & \cellcolor[HTML]{4D9573} $0.4$ & \cellcolor[HTML]{4D9573} $0.9$ & \cellcolor[HTML]{E4F4AE} $3.5$\\ 
GW200219\_094415 & \cellcolor[HTML]{4D9573} $0.8$ & \cellcolor[HTML]{4D9573} $1.0$ & \cellcolor[HTML]{FEFFD1} $4.0$\\ 
GW200220\_124850 & \cellcolor[HTML]{4D9573} $0.2$ & \cellcolor[HTML]{4D9573} $0.3$ & \cellcolor[HTML]{4D9573} $0.7$\\ 
GW200224\_222234 & \cellcolor[HTML]{4D9573} $0.7$ & \cellcolor[HTML]{4D9573} $1.4$ & \cellcolor[HTML]{C04D67} $14.5$\\ 
GW200311\_115853 & \cellcolor[HTML]{4D9573} $1.6$ & \cellcolor[HTML]{87CB8F} $2.6$ & \cellcolor[HTML]{C04D67} $10.2$\\ 
\hline

  \end{tabular}

  \caption{37 binary black hole events from GWTC-2 and GWTC-3 analyzed with \textsc{Dingo} trained with three different PSD datasets. We report the JSD (in units of $10^{-3}$~nat) between \textsc{Dingo} and reference posteriors, averaged over all inferred source parameters.}
  \label{tab:events_JSD}
\end{table*}

\begin{table*}[]
  \centering
  \begin{tabular}{lllr}\hline\hline
        & \multicolumn{3}{c}{\bf{Noise datasets}} \\
         Event & Oracle & Synthetic & Naive\\\hline\hline
GW190408\_181802 & \cellcolor[HTML]{7FC78D} $28.5$ & \cellcolor[HTML]{D2ECA2} $19.4$ & \cellcolor[HTML]{E5F4AF} $17.3$\\ 
GW190413\_052954 & \cellcolor[HTML]{4D9573} $44.6$ & \cellcolor[HTML]{4D9573} $45.0$ & \cellcolor[HTML]{4D9573} $41.9$\\ 
GW190413\_134308 & \cellcolor[HTML]{4D9573} $41.0$ & \cellcolor[HTML]{4D9573} $49.3$ & \cellcolor[HTML]{4D9573} $43.6$\\ 
GW190421\_213856 & \cellcolor[HTML]{4D9573} $51.1$ & \cellcolor[HTML]{4D9573} $46.2$ & \cellcolor[HTML]{4D9573} $44.7$\\ 
GW190503\_185404 & \cellcolor[HTML]{4D9573} $46.9$ & \cellcolor[HTML]{4D9573} $40.4$ & \cellcolor[HTML]{56A87D} $35.4$\\ 
GW190513\_205428 & \cellcolor[HTML]{81C88D} $28.4$ & \cellcolor[HTML]{83C98E} $28.0$ & \cellcolor[HTML]{F0F9BE} $16.0$\\ 
GW190514\_065416 & \cellcolor[HTML]{59AD7F} $34.5$ & \cellcolor[HTML]{519F78} $37.5$ & \cellcolor[HTML]{52A078} $37.3$\\ 
GW190517\_055101 & \cellcolor[HTML]{C8E89C} $20.5$ & \cellcolor[HTML]{FFFACA} $13.6$ & \cellcolor[HTML]{B2DE95} $23.0$\\ 
GW190519\_153544 & \cellcolor[HTML]{4D9573} $41.0$ & \cellcolor[HTML]{4D9573} $41.7$ & \cellcolor[HTML]{9DD593} $25.3$\\ 
GW190521\_074359 & \cellcolor[HTML]{6ABC87} $30.9$ & \cellcolor[HTML]{85CA8E} $27.8$ & \cellcolor[HTML]{D6EEA5} $18.9$\\ 
GW190527\_092055 & \cellcolor[HTML]{97D292} $25.8$ & \cellcolor[HTML]{5BB181} $33.6$ & \cellcolor[HTML]{7FC78D} $28.6$\\ 
GW190602\_175927 & \cellcolor[HTML]{4D9573} $41.7$ & \cellcolor[HTML]{4D9573} $46.0$ & \cellcolor[HTML]{4F9975} $38.8$\\ 
GW190701\_203306 & \cellcolor[HTML]{89CC8F} $27.5$ & \cellcolor[HTML]{5BB181} $33.5$ & \cellcolor[HTML]{94D192} $26.4$\\ 
GW190719\_215514 & \cellcolor[HTML]{A5D894} $24.3$ & \cellcolor[HTML]{B2DE95} $22.9$ & \cellcolor[HTML]{ACDC94} $23.4$\\ 
GW190727\_060333 & \cellcolor[HTML]{4D9573} $42.7$ & \cellcolor[HTML]{4D9573} $48.7$ & \cellcolor[HTML]{68BB87} $31.1$\\ 
GW190731\_140936 & \cellcolor[HTML]{4D9573} $46.5$ & \cellcolor[HTML]{4D9573} $47.7$ & \cellcolor[HTML]{5BB081} $33.9$\\ 
GW190803\_022701 & \cellcolor[HTML]{4D9573} $51.8$ & \cellcolor[HTML]{4D9573} $45.5$ & \cellcolor[HTML]{4D9573} $48.0$\\ 
GW190805\_211137 & \cellcolor[HTML]{4D9573} $50.1$ & \cellcolor[HTML]{4D9573} $51.8$ & \cellcolor[HTML]{4E9874} $39.2$\\ 
GW190828\_063405 & \cellcolor[HTML]{4D9573} $50.8$ & \cellcolor[HTML]{56A87D} $35.4$ & \cellcolor[HTML]{5AAE80} $34.2$\\ 

\hline

  \end{tabular}
  \hspace{0.5cm}
  \begin{tabular}{lllr}\hline\hline
        & \multicolumn{3}{c}{\bf{Noise datasets}} \\
         Event & Oracle & Synthetic & Naive\\\hline\hline

GW190909\_114149 & \cellcolor[HTML]{5DB483} $32.9$ & \cellcolor[HTML]{E3F3AD} $17.6$ & \cellcolor[HTML]{B5DF95} $22.5$\\ 
GW190915\_235702 & \cellcolor[HTML]{4D9573} $49.3$ & \cellcolor[HTML]{4D9573} $48.7$ & \cellcolor[HTML]{4D9573} $45.1$\\ 
GW190926\_050336 & \cellcolor[HTML]{6CBE88} $30.8$ & \cellcolor[HTML]{B9E196} $22.1$ & \cellcolor[HTML]{87CB8F} $27.7$\\ \hline
\multicolumn{1}{c}{\bf{O3b} $\downarrow$} & & & 
\\\hline
GW191127\_050227 & \cellcolor[HTML]{C04D67} $2.8$ & \cellcolor[HTML]{C04D67} $3.9$ & \cellcolor[HTML]{C04D67} $2.5$\\ 
GW191204\_110529 & \cellcolor[HTML]{D2ECA2} $19.3$ & \cellcolor[HTML]{CDEA9E} $20.1$ & \cellcolor[HTML]{E6F5B1} $17.1$\\ 
GW191215\_223052 & \cellcolor[HTML]{4D9573} $45.0$ & \cellcolor[HTML]{4D9573} $44.1$ & \cellcolor[HTML]{7BC58C} $28.9$\\ 
GW191222\_033537 & \cellcolor[HTML]{4D9573} $46.6$ & \cellcolor[HTML]{4D9573} $47.6$ & \cellcolor[HTML]{4D9774} $39.7$\\ 
GW191230\_180458 & \cellcolor[HTML]{4D9573} $41.3$ & \cellcolor[HTML]{4D9573} $40.3$ & \cellcolor[HTML]{4D9573} $40.5$\\ 
GW200128\_022011 & \cellcolor[HTML]{5BB081} $33.9$ & \cellcolor[HTML]{75C28A} $29.8$ & \cellcolor[HTML]{75C28A} $29.8$\\ 
GW200129\_065458 & \cellcolor[HTML]{89CC8F} $27.4$ & \cellcolor[HTML]{6ABC87} $30.9$ & \cellcolor[HTML]{FAAB84} $8.2$\\ 
GW200208\_130117 & \cellcolor[HTML]{4D9573} $48.7$ & \cellcolor[HTML]{4D9573} $47.8$ & \cellcolor[HTML]{4D9573} $41.7$\\ 
GW200208\_222617 & \cellcolor[HTML]{FECA93} $9.5$ & \cellcolor[HTML]{F3FAC2} $15.5$ & \cellcolor[HTML]{FFFED0} $13.9$\\ 
GW200209\_085452 & \cellcolor[HTML]{EAF6B5} $16.7$ & \cellcolor[HTML]{FFF7C5} $13.1$ & \cellcolor[HTML]{EBF6B6} $16.6$\\ 
GW200216\_220804 & \cellcolor[HTML]{5BB081} $33.7$ & \cellcolor[HTML]{5EB684} $32.5$ & \cellcolor[HTML]{6EBF88} $30.6$\\ 
GW200219\_094415 & \cellcolor[HTML]{9DD593} $25.4$ & \cellcolor[HTML]{B5DF95} $22.6$ & \cellcolor[HTML]{AEDC95} $23.3$\\ 
GW200220\_124850 & \cellcolor[HTML]{4D9573} $49.9$ & \cellcolor[HTML]{4D9573} $47.7$ & \cellcolor[HTML]{4D9573} $46.6$\\ 
GW200224\_222234 & \cellcolor[HTML]{4D9573} $49.5$ & \cellcolor[HTML]{55A57B} $36.0$ & \cellcolor[HTML]{FEC892} $9.5$\\ 
GW200311\_115853 & \cellcolor[HTML]{4D9573} $43.9$ & \cellcolor[HTML]{4D9573} $41.8$ & \cellcolor[HTML]{EDF7B9} $16.3$\\ 

\hline

  \end{tabular}

  \caption{Importance sampling efficiency $\epsilon$ for \textsc{Dingo} networks trained with three different PSD datasets, based on 100,000 samples per analysis. Since the variance of $\epsilon$ between identical importance sampling runs can be high, we here report median scores over 10 runs for each event.}
  \label{tab:events_ESS_median}
\end{table*}

\subsection{Real data}

We now perform a large study on real data~\citep{Abbott:2019ebz}, analyzing all 37 BBH events from GWTC-2 and GWTC-3 that are consistent with the prior.
Here, we use four different \textsc{Dingo} models with distance priors $[0.1,2]$~Gpc, $[0.1,4]$~Gpc, $[0.1,6]$~Gpc and $[0.1,12]$~Gpc.
The mean JSDs over all 15 parameters are reported in Table \ref{tab:events_JSD}.
Generally, the scores fluctuate more compared to the highly-controlled simulated data, in that the noise realizations vary and can be non-stationary or non-Gaussian, and real signals do not precisely match models. GW190527\_092055, for example, is more difficult for the \textit{Oracle} network even though the PSD is covered by the training dataset.
We see that the trend observed for simulated data translates to real events and all 15 parameters. The decreasing accuracy for the \textit{Naive} baseline becomes particularly apparent after the transition from O3a to O3b. With our proposed \textit{Synthetic} noise model, however, the accuracy is maintained throughout the entire observing run, and on par with the \textit{Oracle} performance. This showcases, once again, that our approach is indeed capable of forecasting shifts in the PSD distribution and to enable robust inference.

GW190517\_055101 has a high mean JSD for all three datasets, suggesting that the poor performance is not due to an out-of-distribution PSD.
Across all events (except GW190517\_055101) and parameters, we obtain average JSDs of $1.2 \cdot 10^{-3}$~nat for the \emph{Oracle} noise dataset, $1.4 \cdot 10^{-3}$~nat with the \emph{Synthetic} noise dataset and $2.7 \cdot 10^{-3}$~nat with the \emph{Naive} noise dataset. We thus conclude that our approach serves as a convincing replacement to the empirical PSD distribution when full PSD information is unavailable.

We report the importance sampling efficiency for all analyzed events in Table \ref{tab:events_ESS_median}. We see that high overall scores are achieved with all PSD datasets, although scores are usually lowest with the \emph{Naive} noise dataset.
Compared to the simulated events of Section~\ref{sec:sim-data}, the downward trend in sampling efficiency for the \emph{Naive} noise dataset is somewhat less clear. This is likely due to varying noise realizations and source parameters across real events, which introduce additional complications when comparing sampling efficiencies.
For visual comparison, we include corner plots of selected events in the appendix.

\subsection{Discussion}
We compared the performance of \textsc{Dingo} inference networks trained with empirical and synthetic PSD distributions. In training, these distributions are represented as a finite set of PSD samples, and the empirical distributions consist of far fewer samples per detector (\textit{Naive}: 100 PSDs, \textit{Oracle}: $\sim$ 5,000 PSDs) than the \textit{Synthetic} distribution (50,000 PSDs). The unbalanced nature of these datasets may impact our results, since more training samples naturally lead to better generalization. However, the effectively unlimited number of synthetic samples available should be viewed as an advantage, and in practice, the number of available samples will always depend on the estimation method. Indeed, for the \textit{Naive} dataset, which consists of PSDs from the first four days of O3, the data scarcity severely limits the number of available PSDs. Therefore, using unbalanced datasets correctly reflects the practical considerations. Generally, however, our results showing a time-dependent performance decay for the \emph{Naive} dataset indicate that the distribution shift has a far more significant effect on performance than the size of the dataset.

\section{Conclusions}

We developed a probabilistic model for detector noise PSDs to improve training for flow-based GW inference and enable low-latency analyses. The PSD model can be fitted to empirical distributions and then rapidly sampled to obtain synthetic PSDs. By design, the PSD model operates on an interpretable latent space, such that samples can be modified in a physically meaningful way. This allows for data-efficient modelling of distribution shifts, as may occur between LVK observing runs.

In our experiments, we fitted a PSD distribution based on data from the second observing run (O2), and predicted the updated distribution for O3 based on only a single PSD from the beginning of O3. We demonstrated on simulated and real GW events that \textsc{Dingo} models trained with these synthetic O3 PSDs perform accurate inference throughout the entire run ($\sim$ 1 year). We found comparable performance to \textsc{Dingo} models that have direct access to the entire O3 PSD distribution. This is in stark contrast to \textsc{Dingo} models trained with PSDs from the first few days of O3, for which the accuracy strongly degrades over time. 
The code used for the results in this paper will soon be made publically available as part of the \textsc{Dingo} Python package.

Our approach crucially hinges on the \emph{conditioning} of \textsc{Dingo} models on the PSD. Indeed, we do not need to predict specific distribution shifts; instead, it is merely necessary that PSDs encountered at inference time lie within the \emph{support} of the training distribution. Anticipated distribution shifts are addressed in two ways: first, the broad-band noise is shifted to better match the target; second, the PSD distribution is artificially broadened to account for unknown (future) variations. Our synthetic PSD distribution is therefore likely broader than necessary (which may result in slightly more difficult training of \textsc{Dingo} networks), but it captures unseen distribution shifts.

These systematic interventions on the PSD distribution are enabled by the design of our PSD model. We disentangle independent factors of variations by separating spectral lines from the broad-band noise, both of which are represented in a latent space.

Our framework further provides a tractable evidence of PSDs under the estimated distribution, which can be used to quantify distribution shifts.
Going forward, this metric could be used to verify that our PSD model continues to work well between future observing runs, e.g.\ O3 and O4. A low evidence would then flag a distribution shift and indicate that re-training the network is required. 
Moreover, for individual events, results can be validated using importance sampling~\cite{Dax:2022pxd}.

Techniques for parametrizing and estimating PSDs such as \textsc{BayesWave}~\cite{Cornish:2014kda} and \textsc{BayesLine}~\cite{Cornish:2014bl} have been successfully applied to spectral estimation and accurate posterior inference, and our method indeed draws inspiration from these. However, our goal is to model a \emph{distribution} of PSDs, whereas these methods are designed to estimate a \emph{single} PSD. Our work in fact complements these methods: by setting $\sigma^2=0$, we can model smooth broad-band noise, such that our synthetic PSDs closely resemble those of \textsc{BayesWave} (as opposed to noisier Welch PSDs). By training \textsc{Dingo} networks using $\sigma^2=0$ synthetic PSDs, we therefore hope to enable inference with \textsc{BayesWave} PSDs as context.

Our PSD model represents a key component in training flow-based inference networks for low-latency analysis of GW data---enabling complete parameter estimation in real time, with no retraining even for unseen PSD distribution shifts (as expected during an observing run). Once \textsc{Dingo} is extended to binary neutron stars, this will play a crucial role in delivering rapid and accurate multimessenger alerts.

\begin{acknowledgments}
  We thank N. Gupte, I. Harry, S. Ossokine, V. Raymond and R. Smith for useful discussions.
  This material is based upon work supported by NSF's
  LIGO Laboratory which is a major facility fully funded by the
  National Science Foundation. This research has made use of data or
  software obtained from the Gravitational Wave Open Science Center
  (gw-openscience.org), a service of LIGO Laboratory, the LIGO
  Scientific Collaboration, the Virgo Collaboration, and KAGRA. LIGO
  Laboratory and Advanced LIGO are funded by the United States
  National Science Foundation (NSF) as well as the Science and
  Technology Facilities Council (STFC) of the United Kingdom, the
  Max-Planck-Society (MPS), and the State of Niedersachsen/Germany for
  support of the construction of Advanced LIGO and construction and
  operation of the GEO600 detector. Additional support for Advanced
  LIGO was provided by the Australian Research Council. Virgo is
  funded, through the European Gravitational Observatory (EGO), by the
  French Centre National de Recherche Scientifique (CNRS), the Italian
  Istituto Nazionale di Fisica Nucleare (INFN) and the Dutch Nikhef,
  with contributions by institutions from Belgium, Germany, Greece,
  Hungary, Ireland, Japan, Monaco, Poland, Portugal, Spain. The
  construction and operation of KAGRA are funded by Ministry of
  Education, Culture, Sports, Science and Technology (MEXT), and Japan
  Society for the Promotion of Science (JSPS), National Research
  Foundation (NRF) and Ministry of Science and ICT (MSIT) in Korea,
  Academia Sinica (AS) and the Ministry of Science and Technology
  (MoST) in Taiwan.  M.D. thanks the Hector Fellow Academy for
  support. J.H.M. and B.S. are members of the MLCoE, EXC number 2064/1
  – Project number 390727645 and the Tübingen AI Center funded by the
  German Ministry for Science and Education (FKZ 01IS18039A).  For the
  implementation of \textsc{Dingo} we use
  \verb|PyTorch|~\cite{NEURIPS2019_9015}, \verb|nflows|~\cite{nflows},
  \verb|LALSimulation|~\cite{lalsuite} and the adam
  optimizer~\cite{Kingma:2014vow}. The plots are generated with
  \verb|matplotlib|~\cite{Hunter:2007} and
  \verb|ChainConsumer|~\cite{Hinton2016}.
\end{acknowledgments}

\appendix

\begin{figure}
\centering
\includegraphics[width=.48\textwidth]{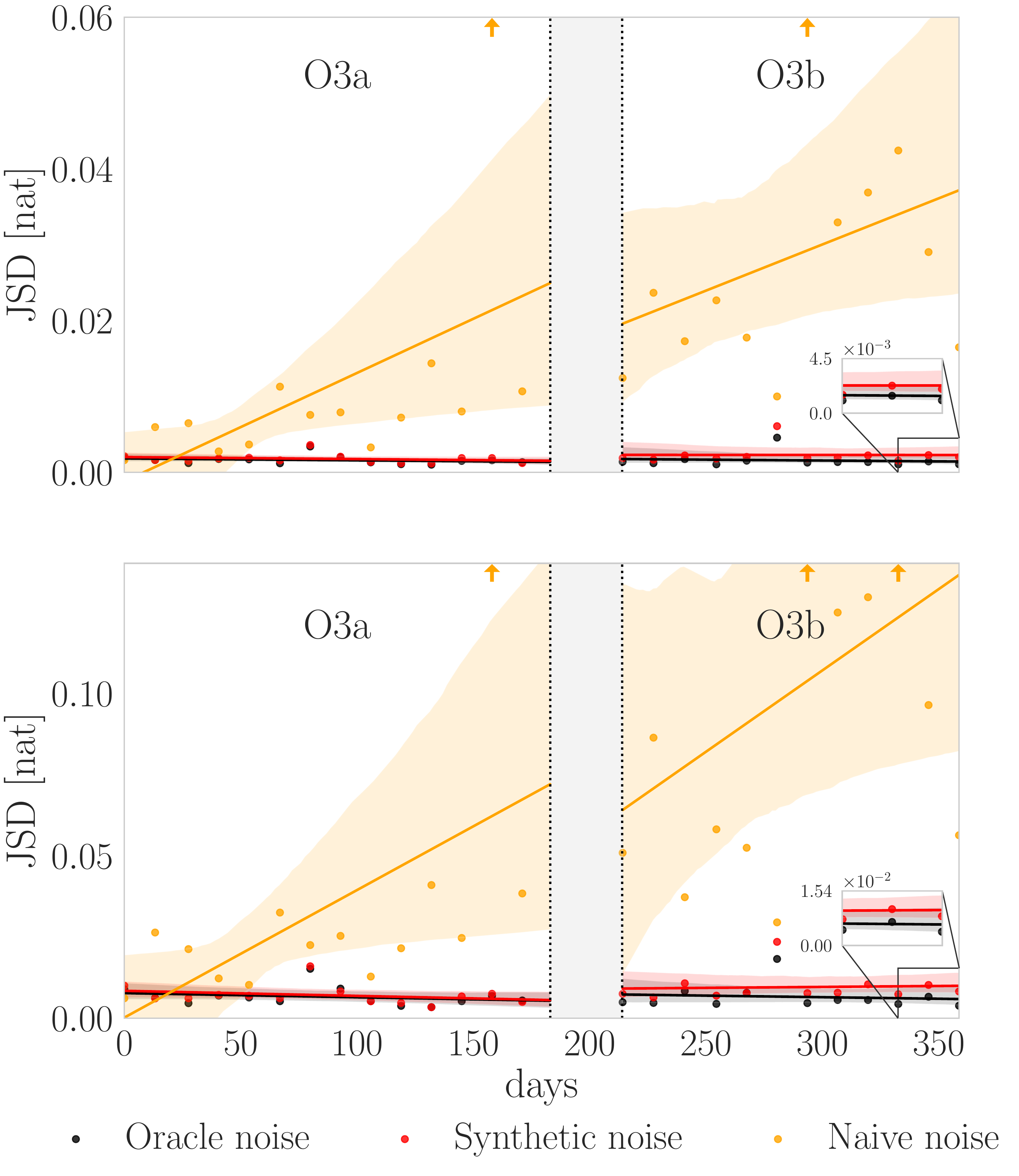}

\caption{Average (top) and maximum (bottom) JS-divergences over all parameters between \textsc{Dingo} samples and reference samples. Injections use noise PSDs estimated at various times throughout O3 (indicated on horizontal axis). Results are averaged over injections with five random sets of source parameters (fixed for all PSDs). Day 0 marks the beginning of O3a. The gap indicates the period when detectors were offline between O3a and O3b. 
}
\label{fig:Avg_JSD_all_params}
\end{figure}

\section{Additional Results}

\begin{figure*} 
  \centering
  \includegraphics[width=0.35\textwidth]{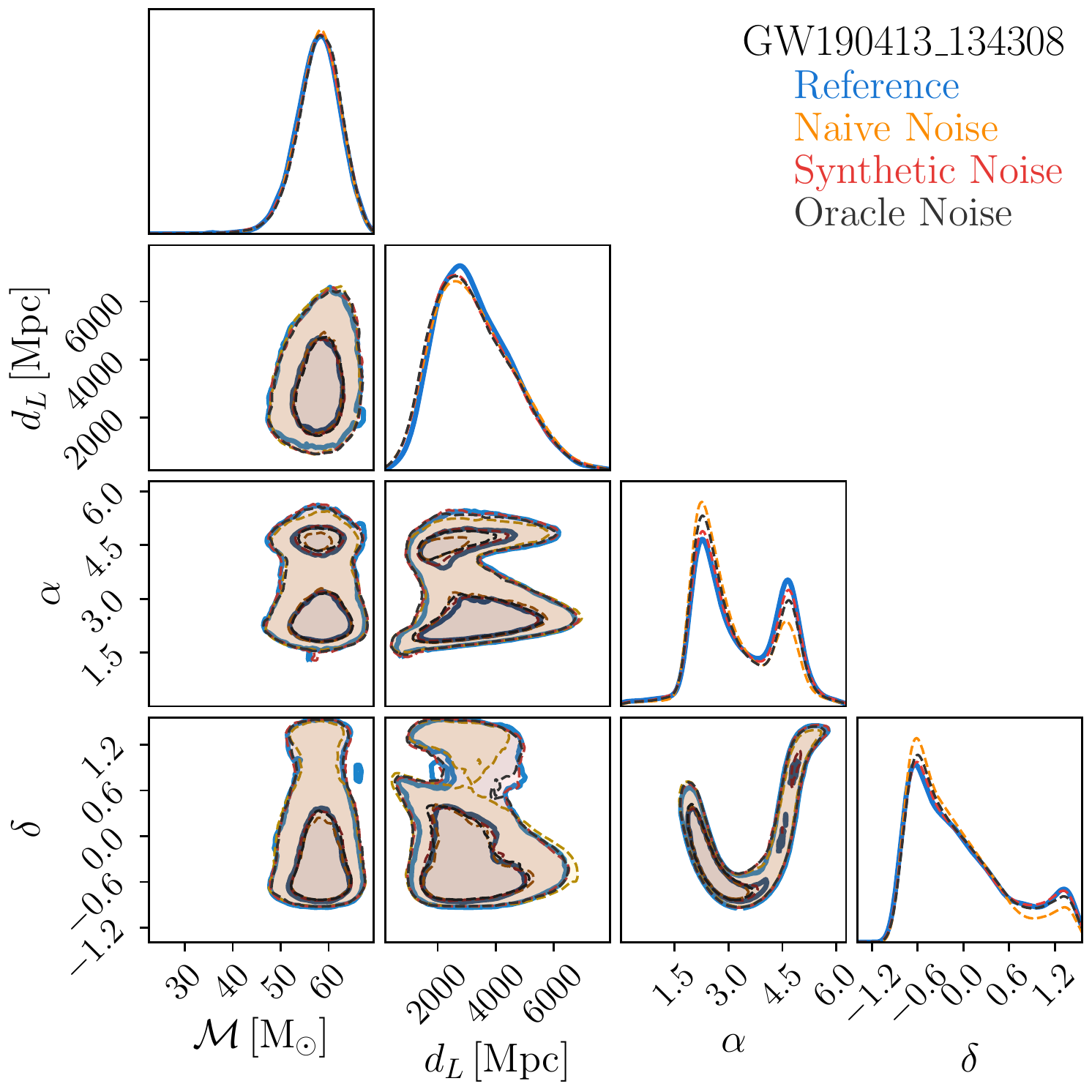}
  \hspace{1.3cm}
  \includegraphics[width=0.35\textwidth]{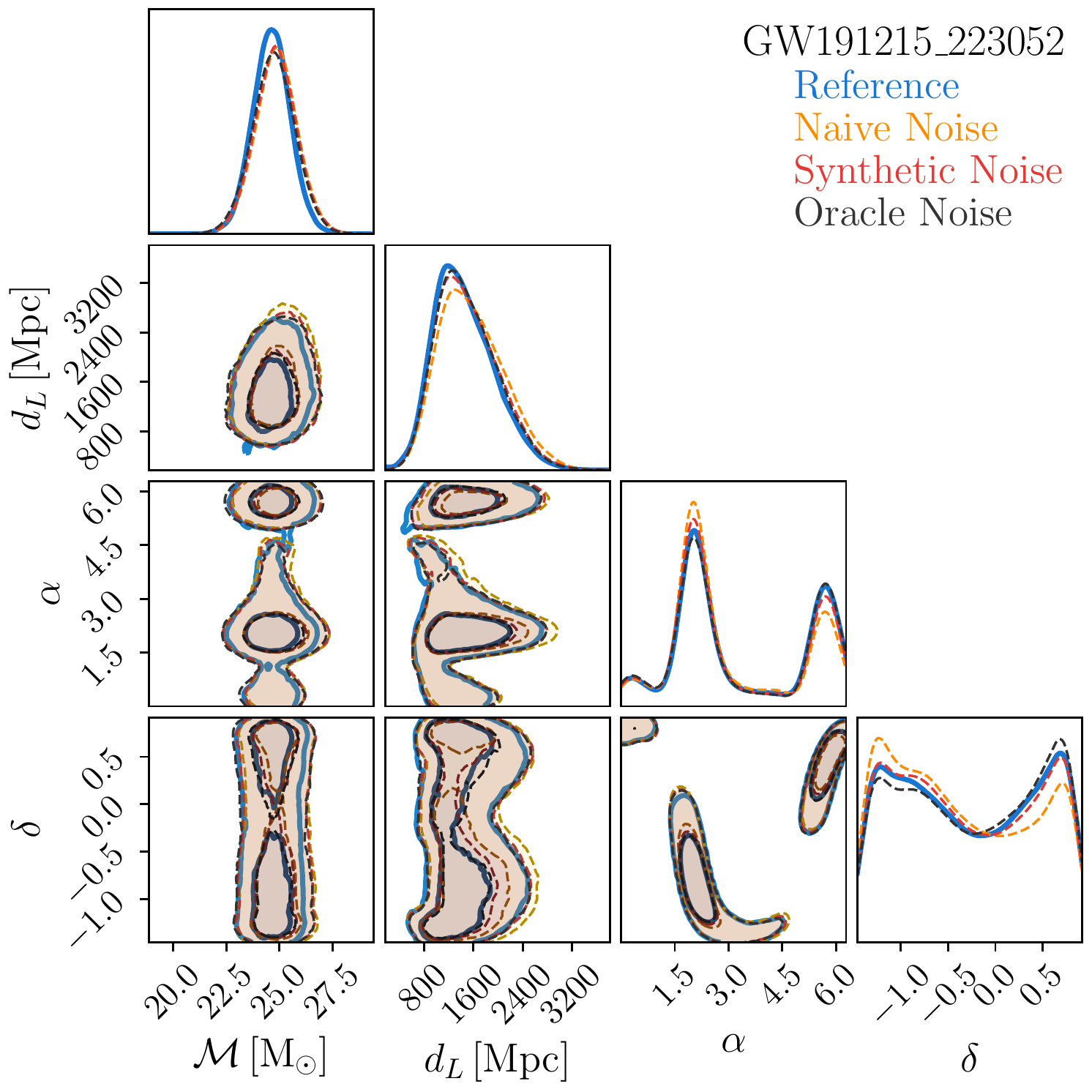}\\
  \includegraphics[width=0.35\textwidth]{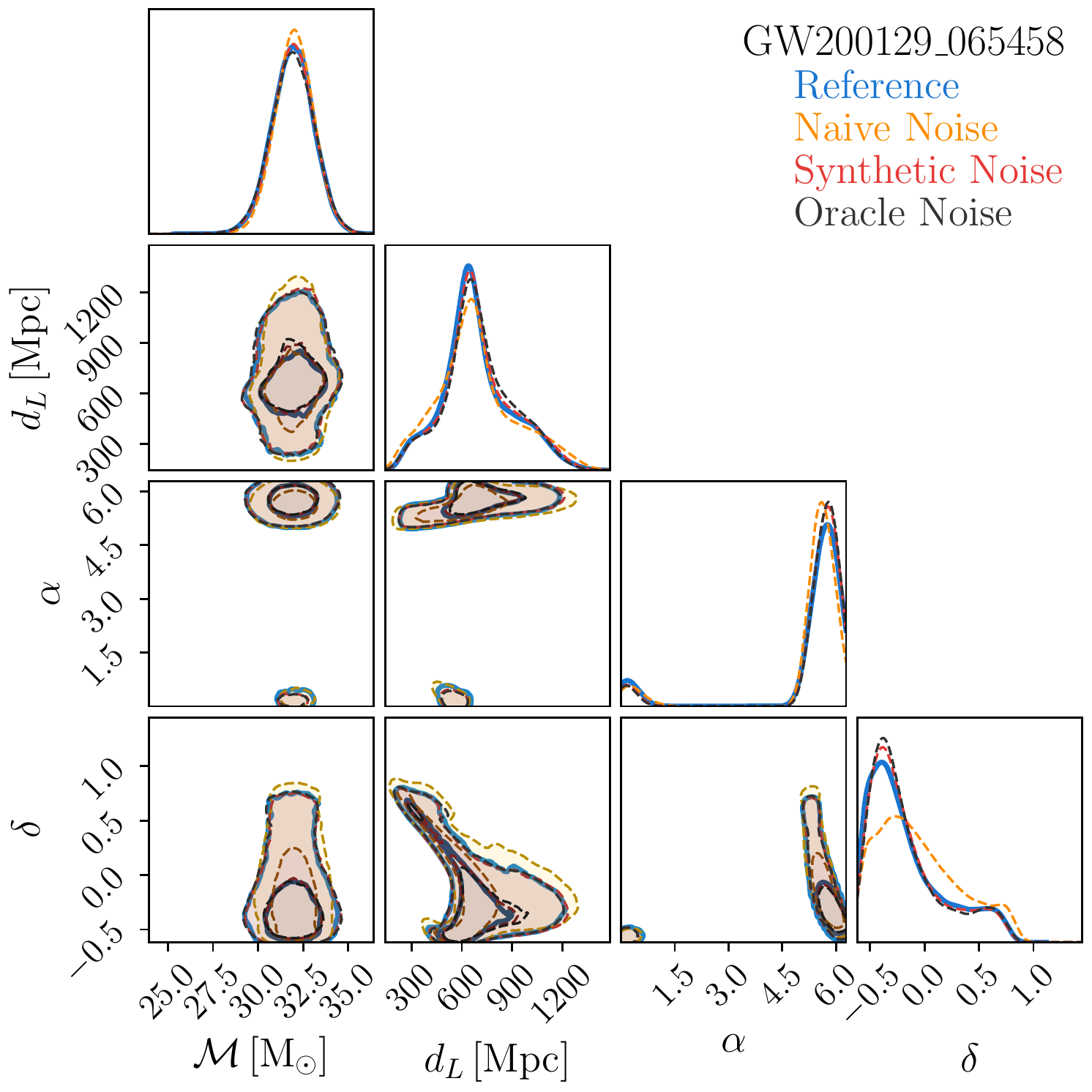}
  \hspace{1.3cm}
  \includegraphics[width=0.35\textwidth]{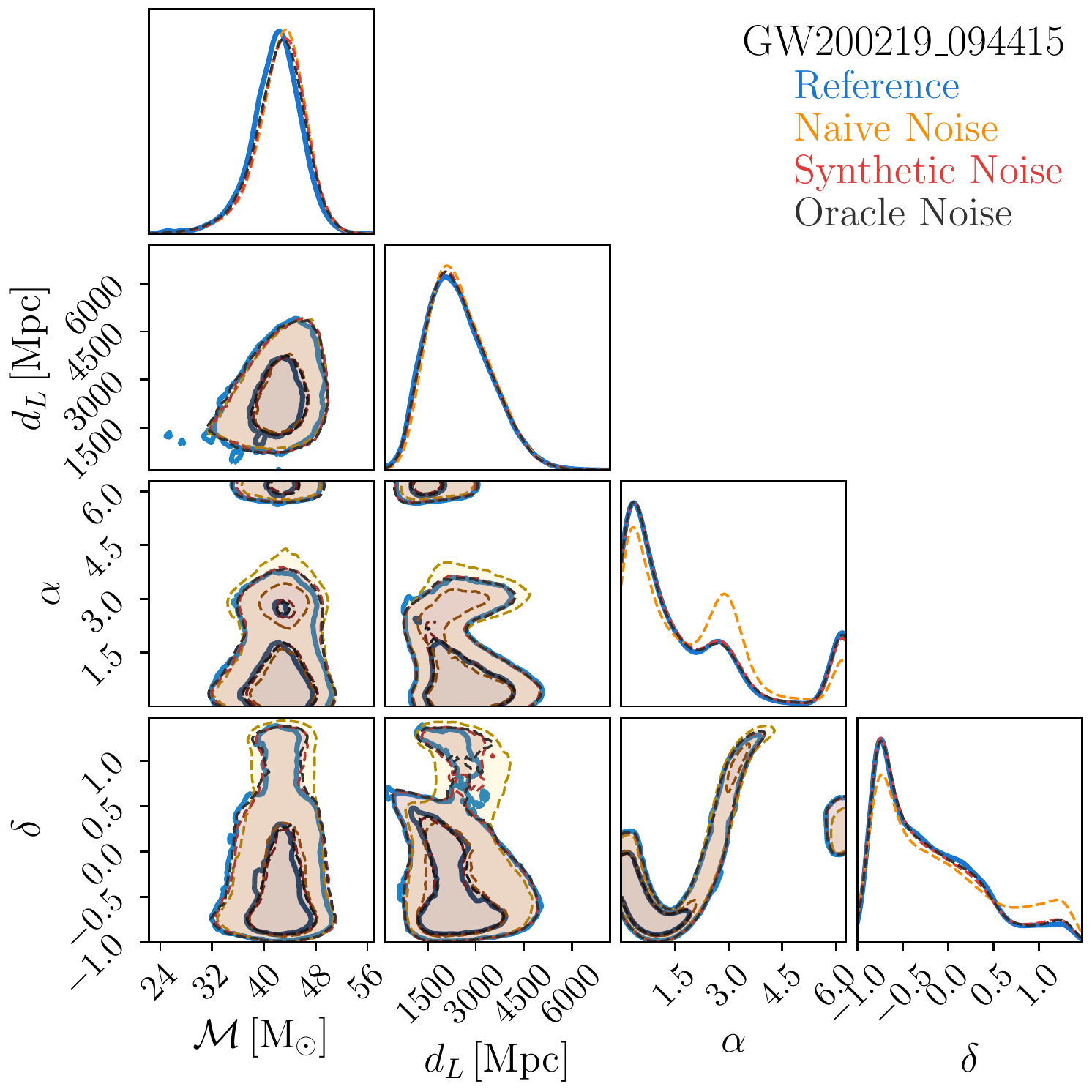}\\
  \includegraphics[width=0.35\textwidth]{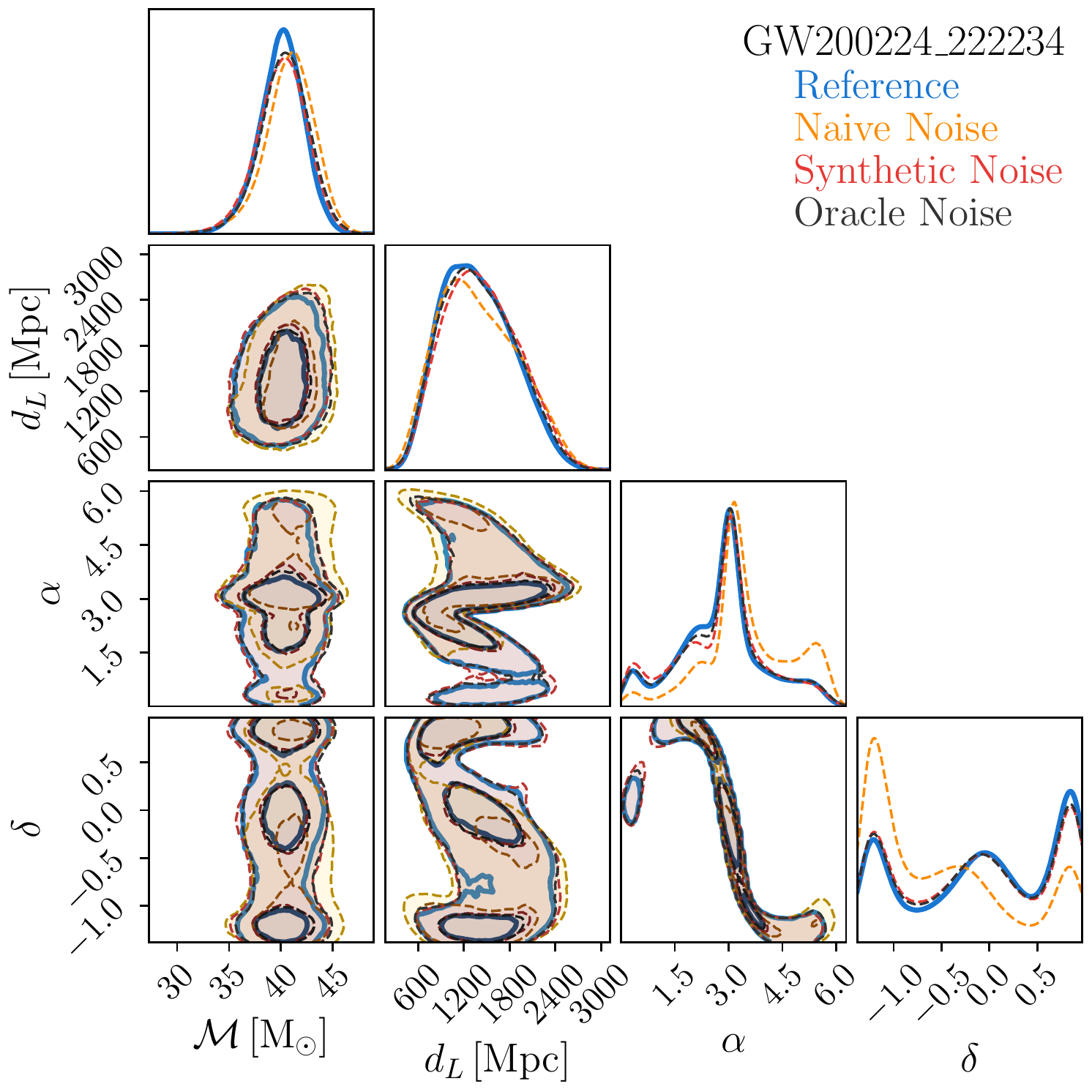}
  \hspace{1.3cm}
  \includegraphics[width=0.35\textwidth]{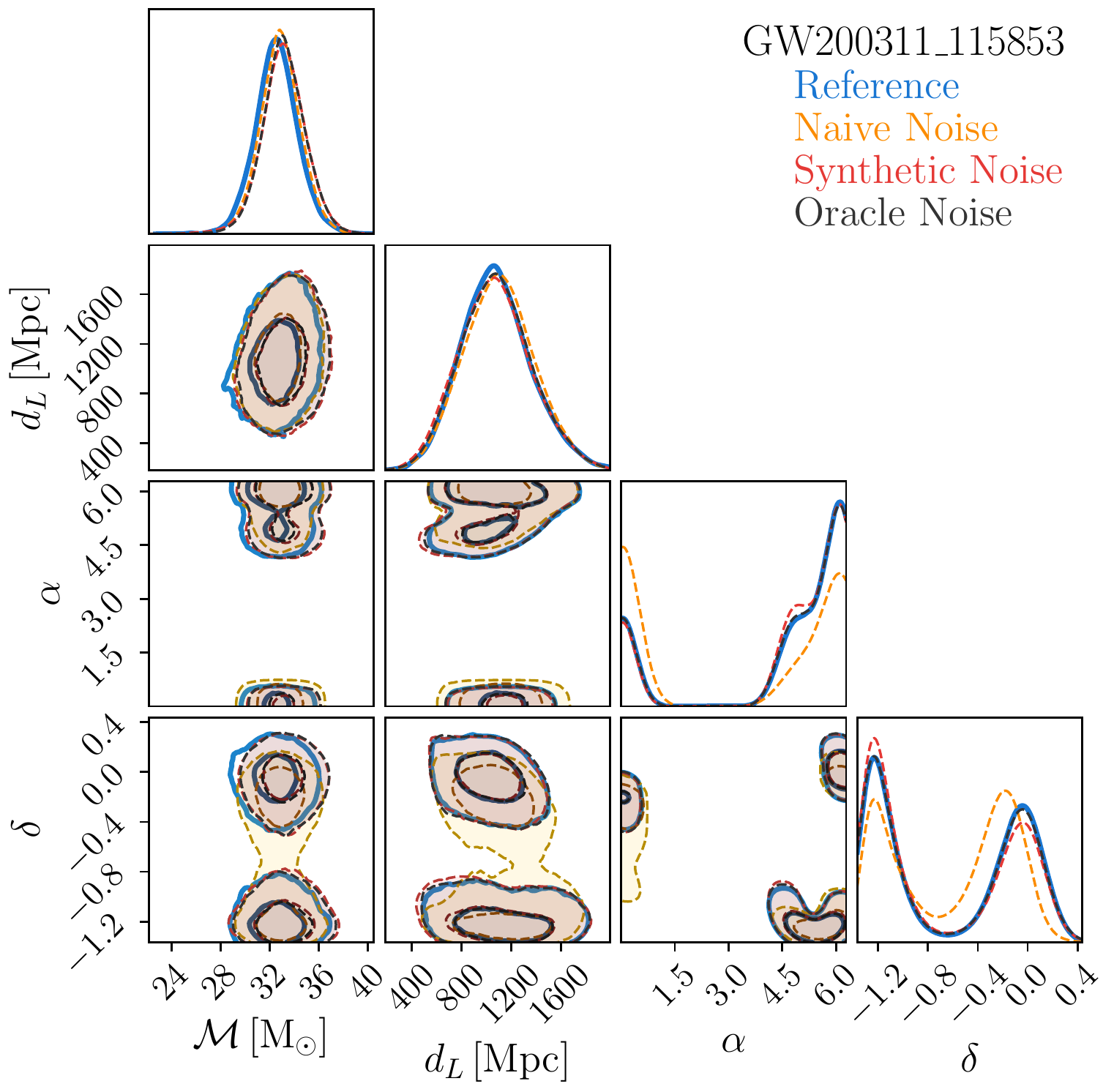}
  \caption{\label{fig:O3_cornerplots}Marginalized one- and two-dimensional posterior distributions for selected O3 events, comparing \textsc{Dingo} models trained with empirically estimated PSDs from the beginning of the observing run (orange), throughout the entire observing run (black) and our proposed synthetic PSD model (red) to the reference posterior (blue, obtained with importance sampling \citep{Dax:2022pxd}). Contours represent 90\% credible regions.
  }
  \label{fig:cornerplots}
\end{figure*}

Figure \ref{fig:Avg_JSD_all_params} shows the mean (top) and maximum (bottom) JSDs between various \textsc{Dingo} posteriors and reference samples, across all inferred parameters, for the study on simulated data in Section \ref{sec:sim-data}. We see that the network trained with \emph{Synthetic} PSDs on average performs similarly to that trained with full PSD information (\emph{Oracle}), with minor deviations for those parameters that are estimated the worst. The average JSD for the \emph{Oracle} dataset is $1.6 \pm 0.7$ $(\cdot 10^{-3})$~nat, for \emph{Synthetic} is $2.0 \pm 0.9$ $(\cdot 10^{-3})$~nat and for \emph{Naive} is
$19.1 \pm 18.8$ $(\cdot 10^{-3})$~nat. 
In Figure \ref{fig:cornerplots}, we compare posterior marginal distributions of selected O3 events between the reference posterior and our \textsc{Dingo} models. With the \emph{Oracle} and \emph{Synthetic} PSD distributions, we obtain excellent agreement with the reference. Using the \emph{Naive} noise model, however, can lead to significant deviations.

\bibliography{mybib}

\begin{thebibliography}{41}%
\makeatletter
\providecommand \@ifxundefined [1]{%
 \@ifx{#1\undefined}
}%
\providecommand \@ifnum [1]{%
 \ifnum #1\expandafter \@firstoftwo
 \else \expandafter \@secondoftwo
 \fi
}%
\providecommand \@ifx [1]{%
 \ifx #1\expandafter \@firstoftwo
 \else \expandafter \@secondoftwo
 \fi
}%
\providecommand \natexlab [1]{#1}%
\providecommand \enquote  [1]{``#1''}%
\providecommand \bibnamefont  [1]{#1}%
\providecommand \bibfnamefont [1]{#1}%
\providecommand \citenamefont [1]{#1}%
\providecommand \href@noop [0]{\@secondoftwo}%
\providecommand \href [0]{\begingroup \@sanitize@url \@href}%
\providecommand \@href[1]{\@@startlink{#1}\@@href}%
\providecommand \@@href[1]{\endgroup#1\@@endlink}%
\providecommand \@sanitize@url [0]{\catcode `\\12\catcode `\$12\catcode
  `\&12\catcode `\#12\catcode `\^12\catcode `\_12\catcode `\%12\relax}%
\providecommand \@@startlink[1]{}%
\providecommand \@@endlink[0]{}%
\providecommand \url  [0]{\begingroup\@sanitize@url \@url }%
\providecommand \@url [1]{\endgroup\@href {#1}{\urlprefix }}%
\providecommand \urlprefix  [0]{URL }%
\providecommand \Eprint [0]{\href }%
\providecommand \doibase [0]{https://doi.org/}%
\providecommand \selectlanguage [0]{\@gobble}%
\providecommand \bibinfo  [0]{\@secondoftwo}%
\providecommand \bibfield  [0]{\@secondoftwo}%
\providecommand \translation [1]{[#1]}%
\providecommand \BibitemOpen [0]{}%
\providecommand \bibitemStop [0]{}%
\providecommand \bibitemNoStop [0]{.\EOS\space}%
\providecommand \EOS [0]{\spacefactor3000\relax}%
\providecommand \BibitemShut  [1]{\csname bibitem#1\endcsname}%
\let\auto@bib@innerbib\@empty
\bibitem [{\citenamefont {Aasi}\ \emph {et~al.}(2015)\citenamefont {Aasi} \emph
  {et~al.}}]{TheLIGOScientific:2014jea}%
  \BibitemOpen
  \bibfield  {author} {\bibinfo {author} {\bibfnamefont {J.}~\bibnamefont
  {Aasi}} \emph {et~al.} (\bibinfo {collaboration} {LIGO Scientific}),\
  }\bibfield  {title} {\bibinfo {title} {{Advanced LIGO}},\ }\href
  {https://doi.org/10.1088/0264-9381/32/7/074001} {\bibfield  {journal}
  {\bibinfo  {journal} {Class. Quant. Grav.}\ }\textbf {\bibinfo {volume}
  {32}},\ \bibinfo {pages} {074001} (\bibinfo {year} {2015})},\ \Eprint
  {https://arxiv.org/abs/1411.4547} {arXiv:1411.4547 [gr-qc]} \BibitemShut
  {NoStop}%
\bibitem [{\citenamefont {Acernese}\ \emph {et~al.}(2015)\citenamefont
  {Acernese} \emph {et~al.}}]{TheVirgo:2014hva}%
  \BibitemOpen
  \bibfield  {author} {\bibinfo {author} {\bibfnamefont {F.}~\bibnamefont
  {Acernese}} \emph {et~al.} (\bibinfo {collaboration} {VIRGO}),\ }\bibfield
  {title} {\bibinfo {title} {{Advanced Virgo: a second-generation
  interferometric gravitational wave detector}},\ }\href
  {https://doi.org/10.1088/0264-9381/32/2/024001} {\bibfield  {journal}
  {\bibinfo  {journal} {Class. Quant. Grav.}\ }\textbf {\bibinfo {volume}
  {32}},\ \bibinfo {pages} {024001} (\bibinfo {year} {2015})},\ \Eprint
  {https://arxiv.org/abs/1408.3978} {arXiv:1408.3978 [gr-qc]} \BibitemShut
  {NoStop}%
\bibitem [{\citenamefont {Somiya}(2012)}]{Somiya:2011np}%
  \BibitemOpen
  \bibfield  {author} {\bibinfo {author} {\bibfnamefont {K.}~\bibnamefont
  {Somiya}} (\bibinfo {collaboration} {KAGRA}),\ }\bibfield  {title} {\bibinfo
  {title} {{Detector configuration of KAGRA: The Japanese cryogenic
  gravitational-wave detector}},\ }\href
  {https://doi.org/10.1088/0264-9381/29/12/124007} {\bibfield  {journal}
  {\bibinfo  {journal} {Class. Quant. Grav.}\ }\textbf {\bibinfo {volume}
  {29}},\ \bibinfo {pages} {124007} (\bibinfo {year} {2012})},\ \Eprint
  {https://arxiv.org/abs/1111.7185} {arXiv:1111.7185 [gr-qc]} \BibitemShut
  {NoStop}%
\bibitem [{\citenamefont {Aso}\ \emph {et~al.}(2013)\citenamefont {Aso},
  \citenamefont {Michimura}, \citenamefont {Somiya}, \citenamefont {Ando},
  \citenamefont {Miyakawa}, \citenamefont {Sekiguchi}, \citenamefont
  {Tatsumi},\ and\ \citenamefont {Yamamoto}}]{Aso:2013eba}%
  \BibitemOpen
  \bibfield  {author} {\bibinfo {author} {\bibfnamefont {Y.}~\bibnamefont
  {Aso}}, \bibinfo {author} {\bibfnamefont {Y.}~\bibnamefont {Michimura}},
  \bibinfo {author} {\bibfnamefont {K.}~\bibnamefont {Somiya}}, \bibinfo
  {author} {\bibfnamefont {M.}~\bibnamefont {Ando}}, \bibinfo {author}
  {\bibfnamefont {O.}~\bibnamefont {Miyakawa}}, \bibinfo {author}
  {\bibfnamefont {T.}~\bibnamefont {Sekiguchi}}, \bibinfo {author}
  {\bibfnamefont {D.}~\bibnamefont {Tatsumi}},\ and\ \bibinfo {author}
  {\bibfnamefont {H.}~\bibnamefont {Yamamoto}} (\bibinfo {collaboration}
  {KAGRA}),\ }\bibfield  {title} {\bibinfo {title} {{Interferometer design of
  the KAGRA gravitational wave detector}},\ }\href
  {https://doi.org/10.1103/PhysRevD.88.043007} {\bibfield  {journal} {\bibinfo
  {journal} {Phys. Rev. D}\ }\textbf {\bibinfo {volume} {88}},\ \bibinfo
  {pages} {043007} (\bibinfo {year} {2013})},\ \Eprint
  {https://arxiv.org/abs/1306.6747} {arXiv:1306.6747 [gr-qc]} \BibitemShut
  {NoStop}%
\bibitem [{\citenamefont {Akutsu}\ \emph {et~al.}(2021)\citenamefont {Akutsu}
  \emph {et~al.}}]{KAGRA:2020tym}%
  \BibitemOpen
  \bibfield  {author} {\bibinfo {author} {\bibfnamefont {T.}~\bibnamefont
  {Akutsu}} \emph {et~al.} (\bibinfo {collaboration} {KAGRA}),\ }\bibfield
  {title} {\bibinfo {title} {{Overview of KAGRA: Detector design and
  construction history}},\ }\href {https://doi.org/10.1093/ptep/ptaa125}
  {\bibfield  {journal} {\bibinfo  {journal} {PTEP}\ }\textbf {\bibinfo
  {volume} {2021}},\ \bibinfo {pages} {05A101} (\bibinfo {year} {2021})},\
  \Eprint {https://arxiv.org/abs/2005.05574} {arXiv:2005.05574
  [physics.ins-det]} \BibitemShut {NoStop}%
\bibitem [{\citenamefont {Welch}(1967)}]{welch1967use}%
  \BibitemOpen
  \bibfield  {author} {\bibinfo {author} {\bibfnamefont {P.}~\bibnamefont
  {Welch}},\ }\bibfield  {title} {\bibinfo {title} {The use of fast fourier
  transform for the estimation of power spectra: a method based on time
  averaging over short, modified periodograms},\ }\href@noop {} {\bibfield
  {journal} {\bibinfo  {journal} {IEEE Transactions on audio and
  electroacoustics}\ }\textbf {\bibinfo {volume} {15}},\ \bibinfo {pages} {70}
  (\bibinfo {year} {1967})}\BibitemShut {NoStop}%
\bibitem [{\citenamefont {Cornish}\ and\ \citenamefont
  {Littenberg}(2015)}]{Cornish:2014kda}%
  \BibitemOpen
  \bibfield  {author} {\bibinfo {author} {\bibfnamefont {N.~J.}\ \bibnamefont
  {Cornish}}\ and\ \bibinfo {author} {\bibfnamefont {T.~B.}\ \bibnamefont
  {Littenberg}},\ }\bibfield  {title} {\bibinfo {title} {{BayesWave: Bayesian
  Inference for Gravitational Wave Bursts and Instrument Glitches}},\ }\href
  {https://doi.org/10.1088/0264-9381/32/13/135012} {\bibfield  {journal}
  {\bibinfo  {journal} {Class. Quant. Grav.}\ }\textbf {\bibinfo {volume}
  {32}},\ \bibinfo {pages} {135012} (\bibinfo {year} {2015})},\ \Eprint
  {https://arxiv.org/abs/1410.3835} {arXiv:1410.3835 [gr-qc]} \BibitemShut
  {NoStop}%
\bibitem [{\citenamefont {Cornish}\ \emph {et~al.}(2021)\citenamefont
  {Cornish}, \citenamefont {Littenberg}, \citenamefont {B\'ecsy}, \citenamefont
  {Chatziioannou}, \citenamefont {Clark}, \citenamefont {Ghonge},\ and\
  \citenamefont {Millhouse}}]{Cornish:2020dwh}%
  \BibitemOpen
  \bibfield  {author} {\bibinfo {author} {\bibfnamefont {N.~J.}\ \bibnamefont
  {Cornish}}, \bibinfo {author} {\bibfnamefont {T.~B.}\ \bibnamefont
  {Littenberg}}, \bibinfo {author} {\bibfnamefont {B.}~\bibnamefont {B\'ecsy}},
  \bibinfo {author} {\bibfnamefont {K.}~\bibnamefont {Chatziioannou}}, \bibinfo
  {author} {\bibfnamefont {J.~A.}\ \bibnamefont {Clark}}, \bibinfo {author}
  {\bibfnamefont {S.}~\bibnamefont {Ghonge}},\ and\ \bibinfo {author}
  {\bibfnamefont {M.}~\bibnamefont {Millhouse}},\ }\bibfield  {title} {\bibinfo
  {title} {{BayesWave analysis pipeline in the era of gravitational wave
  observations}},\ }\href {https://doi.org/10.1103/PhysRevD.103.044006}
  {\bibfield  {journal} {\bibinfo  {journal} {Phys. Rev. D}\ }\textbf {\bibinfo
  {volume} {103}},\ \bibinfo {pages} {044006} (\bibinfo {year} {2021})},\
  \Eprint {https://arxiv.org/abs/2011.09494} {arXiv:2011.09494 [gr-qc]}
  \BibitemShut {NoStop}%
\bibitem [{\citenamefont {Veitch}\ \emph {et~al.}(2015)\citenamefont {Veitch}
  \emph {et~al.}}]{Veitch:2014wba}%
  \BibitemOpen
  \bibfield  {author} {\bibinfo {author} {\bibfnamefont {J.}~\bibnamefont
  {Veitch}} \emph {et~al.},\ }\bibfield  {title} {\bibinfo {title} {{Parameter
  estimation for compact binaries with ground-based gravitational-wave
  observations using the LALInference software library}},\ }\href
  {https://doi.org/10.1103/PhysRevD.91.042003} {\bibfield  {journal} {\bibinfo
  {journal} {Phys. Rev.}\ }\textbf {\bibinfo {volume} {D91}},\ \bibinfo {pages}
  {042003} (\bibinfo {year} {2015})},\ \Eprint
  {https://arxiv.org/abs/1409.7215} {arXiv:1409.7215 [gr-qc]} \BibitemShut
  {NoStop}%
\bibitem [{\citenamefont {Ashton}\ \emph {et~al.}(2019)\citenamefont {Ashton}
  \emph {et~al.}}]{Ashton:2018jfp}%
  \BibitemOpen
  \bibfield  {author} {\bibinfo {author} {\bibfnamefont {G.}~\bibnamefont
  {Ashton}} \emph {et~al.},\ }\bibfield  {title} {\bibinfo {title} {{BILBY: A
  user-friendly Bayesian inference library for gravitational-wave astronomy}},\
  }\href {https://doi.org/10.3847/1538-4365/ab06fc} {\bibfield  {journal}
  {\bibinfo  {journal} {Astrophys. J. Suppl.}\ }\textbf {\bibinfo {volume}
  {241}},\ \bibinfo {pages} {27} (\bibinfo {year} {2019})},\ \Eprint
  {https://arxiv.org/abs/1811.02042} {arXiv:1811.02042 [astro-ph.IM]}
  \BibitemShut {NoStop}%
\bibitem [{\citenamefont {Romero-Shaw}\ \emph {et~al.}(2020)\citenamefont
  {Romero-Shaw} \emph {et~al.}}]{Romero-Shaw:2020owr}%
  \BibitemOpen
  \bibfield  {author} {\bibinfo {author} {\bibfnamefont {I.~M.}\ \bibnamefont
  {Romero-Shaw}} \emph {et~al.},\ }\bibfield  {title} {\bibinfo {title}
  {{Bayesian inference for compact binary coalescences with bilby: validation
  and application to the first LIGO\textendash{}Virgo gravitational-wave
  transient catalogue}},\ }\href {https://doi.org/10.1093/mnras/staa2850}
  {\bibfield  {journal} {\bibinfo  {journal} {Mon. Not. Roy. Astron. Soc.}\
  }\textbf {\bibinfo {volume} {499}},\ \bibinfo {pages} {3295} (\bibinfo {year}
  {2020})},\ \Eprint {https://arxiv.org/abs/2006.00714} {arXiv:2006.00714
  [astro-ph.IM]} \BibitemShut {NoStop}%
\bibitem [{\citenamefont {Speagle}(2020)}]{Speagle_2020}%
  \BibitemOpen
  \bibfield  {author} {\bibinfo {author} {\bibfnamefont {J.~S.}\ \bibnamefont
  {Speagle}},\ }\bibfield  {title} {\bibinfo {title} {dynesty: a dynamic nested
  sampling package for estimating bayesian posteriors and evidences},\ }\href
  {https://doi.org/10.1093/mnras/staa278} {\bibfield  {journal} {\bibinfo
  {journal} {Monthly Notices of the Royal Astronomical Society}\ }\textbf
  {\bibinfo {volume} {493}},\ \bibinfo {pages} {3132–3158} (\bibinfo {year}
  {2020})},\ \Eprint {https://arxiv.org/abs/1904.02180} {arXiv:1904.02180
  [astro-ph.IM]} \BibitemShut {NoStop}%
\bibitem [{\citenamefont {Gabbard}\ \emph {et~al.}(2022)\citenamefont
  {Gabbard}, \citenamefont {Messenger}, \citenamefont {Heng}, \citenamefont
  {Tonolini},\ and\ \citenamefont {Murray-Smith}}]{Gabbard:2019rde}%
  \BibitemOpen
  \bibfield  {author} {\bibinfo {author} {\bibfnamefont {H.}~\bibnamefont
  {Gabbard}}, \bibinfo {author} {\bibfnamefont {C.}~\bibnamefont {Messenger}},
  \bibinfo {author} {\bibfnamefont {I.~S.}\ \bibnamefont {Heng}}, \bibinfo
  {author} {\bibfnamefont {F.}~\bibnamefont {Tonolini}},\ and\ \bibinfo
  {author} {\bibfnamefont {R.}~\bibnamefont {Murray-Smith}},\ }\bibfield
  {title} {\bibinfo {title} {{Bayesian parameter estimation using conditional
  variational autoencoders for gravitational-wave astronomy}},\ }\href
  {https://doi.org/10.1038/s41567-021-01425-7} {\bibfield  {journal} {\bibinfo
  {journal} {Nature Phys.}\ }\textbf {\bibinfo {volume} {18}},\ \bibinfo
  {pages} {112} (\bibinfo {year} {2022})},\ \Eprint
  {https://arxiv.org/abs/1909.06296} {arXiv:1909.06296 [astro-ph.IM]}
  \BibitemShut {NoStop}%
\bibitem [{\citenamefont {Chua}\ and\ \citenamefont
  {Vallisneri}(2020)}]{Chua:2019wwt}%
  \BibitemOpen
  \bibfield  {author} {\bibinfo {author} {\bibfnamefont {A.~J.~K.}\
  \bibnamefont {Chua}}\ and\ \bibinfo {author} {\bibfnamefont {M.}~\bibnamefont
  {Vallisneri}},\ }\bibfield  {title} {\bibinfo {title} {{Learning Bayesian
  posteriors with neural networks for gravitational-wave inference}},\ }\href
  {https://doi.org/10.1103/PhysRevLett.124.041102} {\bibfield  {journal}
  {\bibinfo  {journal} {Phys. Rev. Lett.}\ }\textbf {\bibinfo {volume} {124}},\
  \bibinfo {pages} {041102} (\bibinfo {year} {2020})},\ \Eprint
  {https://arxiv.org/abs/1909.05966} {arXiv:1909.05966 [gr-qc]} \BibitemShut
  {NoStop}%
\bibitem [{\citenamefont {Chatterjee}\ \emph {et~al.}(2019)\citenamefont
  {Chatterjee}, \citenamefont {Wen}, \citenamefont {Vinsen}, \citenamefont
  {Kovalam},\ and\ \citenamefont {Datta}}]{Chatterjee:2019gqr}%
  \BibitemOpen
  \bibfield  {author} {\bibinfo {author} {\bibfnamefont {C.}~\bibnamefont
  {Chatterjee}}, \bibinfo {author} {\bibfnamefont {L.}~\bibnamefont {Wen}},
  \bibinfo {author} {\bibfnamefont {K.}~\bibnamefont {Vinsen}}, \bibinfo
  {author} {\bibfnamefont {M.}~\bibnamefont {Kovalam}},\ and\ \bibinfo {author}
  {\bibfnamefont {A.}~\bibnamefont {Datta}},\ }\bibfield  {title} {\bibinfo
  {title} {{Using Deep Learning to Localize Gravitational Wave Sources}},\
  }\href {https://doi.org/10.1103/PhysRevD.100.103025} {\bibfield  {journal}
  {\bibinfo  {journal} {Phys. Rev. D}\ }\textbf {\bibinfo {volume} {100}},\
  \bibinfo {pages} {103025} (\bibinfo {year} {2019})},\ \Eprint
  {https://arxiv.org/abs/1909.06367} {arXiv:1909.06367 [astro-ph.IM]}
  \BibitemShut {NoStop}%
\bibitem [{\citenamefont {Green}\ \emph {et~al.}(2020)\citenamefont {Green},
  \citenamefont {Simpson},\ and\ \citenamefont {Gair}}]{Green:2020hst}%
  \BibitemOpen
  \bibfield  {author} {\bibinfo {author} {\bibfnamefont {S.~R.}\ \bibnamefont
  {Green}}, \bibinfo {author} {\bibfnamefont {C.}~\bibnamefont {Simpson}},\
  and\ \bibinfo {author} {\bibfnamefont {J.}~\bibnamefont {Gair}},\ }\bibfield
  {title} {\bibinfo {title} {{Gravitational-wave parameter estimation with
  autoregressive neural network flows}},\ }\href
  {https://doi.org/10.1103/PhysRevD.102.104057} {\bibfield  {journal} {\bibinfo
   {journal} {Phys. Rev. D}\ }\textbf {\bibinfo {volume} {102}},\ \bibinfo
  {pages} {104057} (\bibinfo {year} {2020})},\ \Eprint
  {https://arxiv.org/abs/2002.07656} {arXiv:2002.07656 [astro-ph.IM]}
  \BibitemShut {NoStop}%
\bibitem [{\citenamefont {Green}\ and\ \citenamefont
  {Gair}(2021)}]{Green:2020dnx}%
  \BibitemOpen
  \bibfield  {author} {\bibinfo {author} {\bibfnamefont {S.~R.}\ \bibnamefont
  {Green}}\ and\ \bibinfo {author} {\bibfnamefont {J.}~\bibnamefont {Gair}},\
  }\bibfield  {title} {\bibinfo {title} {{Complete parameter inference for
  GW150914 using deep learning}},\ }\href
  {https://doi.org/10.1088/2632-2153/abfaed} {\bibfield  {journal} {\bibinfo
  {journal} {Mach. Learn. Sci. Tech.}\ }\textbf {\bibinfo {volume} {2}},\
  \bibinfo {pages} {03LT01} (\bibinfo {year} {2021})},\ \Eprint
  {https://arxiv.org/abs/2008.03312} {arXiv:2008.03312 [astro-ph.IM]}
  \BibitemShut {NoStop}%
\bibitem [{\citenamefont {Delaunoy}\ \emph {et~al.}(2020)\citenamefont
  {Delaunoy}, \citenamefont {Wehenkel}, \citenamefont {Hinderer}, \citenamefont
  {Nissanke}, \citenamefont {Weniger}, \citenamefont {Williamson},\ and\
  \citenamefont {Louppe}}]{Delaunoy:2020zcu}%
  \BibitemOpen
  \bibfield  {author} {\bibinfo {author} {\bibfnamefont {A.}~\bibnamefont
  {Delaunoy}}, \bibinfo {author} {\bibfnamefont {A.}~\bibnamefont {Wehenkel}},
  \bibinfo {author} {\bibfnamefont {T.}~\bibnamefont {Hinderer}}, \bibinfo
  {author} {\bibfnamefont {S.}~\bibnamefont {Nissanke}}, \bibinfo {author}
  {\bibfnamefont {C.}~\bibnamefont {Weniger}}, \bibinfo {author} {\bibfnamefont
  {A.~R.}\ \bibnamefont {Williamson}},\ and\ \bibinfo {author} {\bibfnamefont
  {G.}~\bibnamefont {Louppe}},\ }\bibfield  {title} {\bibinfo {title}
  {{Lightning-Fast Gravitational Wave Parameter Inference through Neural
  Amortization}},\ }\href@noop {} {\  (\bibinfo {year} {2020})},\ \Eprint
  {https://arxiv.org/abs/2010.12931} {arXiv:2010.12931 [astro-ph.IM]}
  \BibitemShut {NoStop}%
\bibitem [{\citenamefont {Dax}\ \emph {et~al.}(2021)\citenamefont {Dax},
  \citenamefont {Green}, \citenamefont {Gair}, \citenamefont {Macke},
  \citenamefont {Buonanno},\ and\ \citenamefont {Sch\"olkopf}}]{Dax:2021tsq}%
  \BibitemOpen
  \bibfield  {author} {\bibinfo {author} {\bibfnamefont {M.}~\bibnamefont
  {Dax}}, \bibinfo {author} {\bibfnamefont {S.~R.}\ \bibnamefont {Green}},
  \bibinfo {author} {\bibfnamefont {J.}~\bibnamefont {Gair}}, \bibinfo {author}
  {\bibfnamefont {J.~H.}\ \bibnamefont {Macke}}, \bibinfo {author}
  {\bibfnamefont {A.}~\bibnamefont {Buonanno}},\ and\ \bibinfo {author}
  {\bibfnamefont {B.}~\bibnamefont {Sch\"olkopf}},\ }\bibfield  {title}
  {\bibinfo {title} {{Real-Time Gravitational Wave Science with Neural
  Posterior Estimation}},\ }\href
  {https://doi.org/10.1103/PhysRevLett.127.241103} {\bibfield  {journal}
  {\bibinfo  {journal} {Phys. Rev. Lett.}\ }\textbf {\bibinfo {volume} {127}},\
  \bibinfo {pages} {241103} (\bibinfo {year} {2021})},\ \Eprint
  {https://arxiv.org/abs/2106.12594} {arXiv:2106.12594 [gr-qc]} \BibitemShut
  {NoStop}%
\bibitem [{\citenamefont {Dax}\ \emph {et~al.}(2022{\natexlab{a}})\citenamefont
  {Dax}, \citenamefont {Green}, \citenamefont {Gair}, \citenamefont {Deistler},
  \citenamefont {Sch\"olkopf},\ and\ \citenamefont {Macke}}]{Dax:2021myb}%
  \BibitemOpen
  \bibfield  {author} {\bibinfo {author} {\bibfnamefont {M.}~\bibnamefont
  {Dax}}, \bibinfo {author} {\bibfnamefont {S.~R.}\ \bibnamefont {Green}},
  \bibinfo {author} {\bibfnamefont {J.}~\bibnamefont {Gair}}, \bibinfo {author}
  {\bibfnamefont {M.}~\bibnamefont {Deistler}}, \bibinfo {author}
  {\bibfnamefont {B.}~\bibnamefont {Sch\"olkopf}},\ and\ \bibinfo {author}
  {\bibfnamefont {J.~H.}\ \bibnamefont {Macke}},\ }\bibfield  {title} {\bibinfo
  {title} {{Group equivariant neural posterior estimation}},\ }in\ \href@noop
  {} {\emph {\bibinfo {booktitle} {International Conference on Learning
  Representations (ICLR 2022)}}}\ (\bibinfo {year} {2022})\ \Eprint
  {https://arxiv.org/abs/2111.13139} {arXiv:2111.13139 [cs.LG]} \BibitemShut
  {NoStop}%
\bibitem [{\citenamefont {Krastev}\ \emph {et~al.}(2021)\citenamefont
  {Krastev}, \citenamefont {Gill}, \citenamefont {Villar},\ and\ \citenamefont
  {Berger}}]{Krastev:2020skk}%
  \BibitemOpen
  \bibfield  {author} {\bibinfo {author} {\bibfnamefont {P.~G.}\ \bibnamefont
  {Krastev}}, \bibinfo {author} {\bibfnamefont {K.}~\bibnamefont {Gill}},
  \bibinfo {author} {\bibfnamefont {V.~A.}\ \bibnamefont {Villar}},\ and\
  \bibinfo {author} {\bibfnamefont {E.}~\bibnamefont {Berger}},\ }\bibfield
  {title} {\bibinfo {title} {{Detection and Parameter Estimation of
  Gravitational Waves from Binary Neutron-Star Mergers in Real LIGO Data using
  Deep Learning}},\ }\href {https://doi.org/10.1016/j.physletb.2021.136161}
  {\bibfield  {journal} {\bibinfo  {journal} {Phys. Lett. B}\ }\textbf
  {\bibinfo {volume} {815}},\ \bibinfo {pages} {136161} (\bibinfo {year}
  {2021})},\ \Eprint {https://arxiv.org/abs/2012.13101} {arXiv:2012.13101
  [astro-ph.IM]} \BibitemShut {NoStop}%
\bibitem [{\citenamefont {Shen}\ \emph {et~al.}(2021)\citenamefont {Shen},
  \citenamefont {Huerta}, \citenamefont {O'Shea}, \citenamefont {Kumar},\ and\
  \citenamefont {Zhao}}]{Shen:2019vep}%
  \BibitemOpen
  \bibfield  {author} {\bibinfo {author} {\bibfnamefont {H.}~\bibnamefont
  {Shen}}, \bibinfo {author} {\bibfnamefont {E.~A.}\ \bibnamefont {Huerta}},
  \bibinfo {author} {\bibfnamefont {E.}~\bibnamefont {O'Shea}}, \bibinfo
  {author} {\bibfnamefont {P.}~\bibnamefont {Kumar}},\ and\ \bibinfo {author}
  {\bibfnamefont {Z.}~\bibnamefont {Zhao}},\ }\bibfield  {title} {\bibinfo
  {title} {{Statistically-informed deep learning for gravitational wave
  parameter estimation}},\ }\href@noop {} {\  (\bibinfo {year} {2021})},\
  \Eprint {https://arxiv.org/abs/1903.01998v3} {arXiv:1903.01998v3 [gr-qc]}
  \BibitemShut {NoStop}%
\bibitem [{\citenamefont {Dax}\ \emph {et~al.}(2022{\natexlab{b}})\citenamefont
  {Dax}, \citenamefont {Green}, \citenamefont {Gair}, \citenamefont {P\"urrer},
  \citenamefont {Wildberger}, \citenamefont {Macke}, \citenamefont {Buonanno},\
  and\ \citenamefont {Sch\"olkopf}}]{Dax:2022pxd}%
  \BibitemOpen
  \bibfield  {author} {\bibinfo {author} {\bibfnamefont {M.}~\bibnamefont
  {Dax}}, \bibinfo {author} {\bibfnamefont {S.~R.}\ \bibnamefont {Green}},
  \bibinfo {author} {\bibfnamefont {J.}~\bibnamefont {Gair}}, \bibinfo {author}
  {\bibfnamefont {M.}~\bibnamefont {P\"urrer}}, \bibinfo {author}
  {\bibfnamefont {J.}~\bibnamefont {Wildberger}}, \bibinfo {author}
  {\bibfnamefont {J.~H.}\ \bibnamefont {Macke}}, \bibinfo {author}
  {\bibfnamefont {A.}~\bibnamefont {Buonanno}},\ and\ \bibinfo {author}
  {\bibfnamefont {B.}~\bibnamefont {Sch\"olkopf}},\ }\bibfield  {title}
  {\bibinfo {title} {{Neural Importance Sampling for Rapid and Reliable
  Gravitational-Wave Inference}},\ }\href@noop {} {\  (\bibinfo {year}
  {2022}{\natexlab{b}})},\ \Eprint {https://arxiv.org/abs/2210.05686}
  {arXiv:2210.05686 [gr-qc]} \BibitemShut {NoStop}%
\bibitem [{\citenamefont {Abbott}\ \emph {et~al.}(2021)\citenamefont {Abbott}
  \emph {et~al.}}]{Abbott:2019ebz}%
  \BibitemOpen
  \bibfield  {author} {\bibinfo {author} {\bibfnamefont {R.}~\bibnamefont
  {Abbott}} \emph {et~al.} (\bibinfo {collaboration} {LIGO Scientific,
  Virgo}),\ }\bibfield  {title} {\bibinfo {title} {{Open data from the first
  and second observing runs of Advanced LIGO and Advanced Virgo}},\ }\href
  {https://doi.org/10.1016/j.softx.2021.100658} {\bibfield  {journal} {\bibinfo
   {journal} {SoftwareX}\ }\textbf {\bibinfo {volume} {13}},\ \bibinfo {pages}
  {100658} (\bibinfo {year} {2021})},\ \Eprint
  {https://arxiv.org/abs/1912.11716} {arXiv:1912.11716 [gr-qc]} \BibitemShut
  {NoStop}%
\bibitem [{\citenamefont {Bartholomew}\ \emph {et~al.}(2011)\citenamefont
  {Bartholomew}, \citenamefont {Knott},\ and\ \citenamefont
  {Moustaki}}]{bartholomew2011latent}%
  \BibitemOpen
  \bibfield  {author} {\bibinfo {author} {\bibfnamefont {D.~J.}\ \bibnamefont
  {Bartholomew}}, \bibinfo {author} {\bibfnamefont {M.}~\bibnamefont {Knott}},\
  and\ \bibinfo {author} {\bibfnamefont {I.}~\bibnamefont {Moustaki}},\
  }\href@noop {} {\emph {\bibinfo {title} {Latent variable models and factor
  analysis: A unified approach}}}\ (\bibinfo  {publisher} {John Wiley \&
  Sons},\ \bibinfo {year} {2011})\BibitemShut {NoStop}%
\bibitem [{\citenamefont {Kingma}\ and\ \citenamefont
  {Welling}(2014)}]{Kingma2014}%
  \BibitemOpen
  \bibfield  {author} {\bibinfo {author} {\bibfnamefont {D.~P.}\ \bibnamefont
  {Kingma}}\ and\ \bibinfo {author} {\bibfnamefont {M.}~\bibnamefont
  {Welling}},\ }\bibfield  {title} {\bibinfo {title} {{Auto-Encoding
  Variational Bayes}},\ }in\ \href@noop {} {\emph {\bibinfo {booktitle} {2nd
  International Conference on Learning Representations, {ICLR} 2014, Banff, AB,
  Canada, April 14-16, 2014, Conference Track Proceedings}}}\ (\bibinfo {year}
  {2014})\ \Eprint {https://arxiv.org/abs/http://arxiv.org/abs/1312.6114v10}
  {http://arxiv.org/abs/1312.6114v10} \BibitemShut {NoStop}%
\bibitem [{\citenamefont {Rezende}\ \emph {et~al.}(2014)\citenamefont
  {Rezende}, \citenamefont {Mohamed},\ and\ \citenamefont
  {Wierstra}}]{rezende2014stochastic}%
  \BibitemOpen
  \bibfield  {author} {\bibinfo {author} {\bibfnamefont {D.~J.}\ \bibnamefont
  {Rezende}}, \bibinfo {author} {\bibfnamefont {S.}~\bibnamefont {Mohamed}},\
  and\ \bibinfo {author} {\bibfnamefont {D.}~\bibnamefont {Wierstra}},\
  }\bibfield  {title} {\bibinfo {title} {Stochastic backpropagation and
  approximate inference in deep generative models},\ }in\ \href@noop {} {\emph
  {\bibinfo {booktitle} {International Conference on Machine Learning}}}\
  (\bibinfo {year} {2014})\ pp.\ \bibinfo {pages} {1278--1286},\ \Eprint
  {https://arxiv.org/abs/1401.4082} {1401.4082 [stat.ML]} \BibitemShut
  {NoStop}%
\bibitem [{\citenamefont {Littenberg}\ and\ \citenamefont
  {Cornish}(2014)}]{Cornish:2014bl}%
  \BibitemOpen
  \bibfield  {author} {\bibinfo {author} {\bibfnamefont {T.}~\bibnamefont
  {Littenberg}}\ and\ \bibinfo {author} {\bibfnamefont {N.}~\bibnamefont
  {Cornish}},\ }\bibfield  {title} {\bibinfo {title} {Bayesline: Bayesian
  inference for spectral estimation of gravitational wave detector noise},\
  }\href {https://doi.org/10.1103/PhysRevD.91.084034} {\bibfield  {journal}
  {\bibinfo  {journal} {Physical Review D}\ }\textbf {\bibinfo {volume} {91}}
  (\bibinfo {year} {2014})}\BibitemShut {NoStop}%
\bibitem [{\citenamefont {Khan}\ \emph {et~al.}(2016)\citenamefont {Khan},
  \citenamefont {Husa}, \citenamefont {Hannam}, \citenamefont {Ohme},
  \citenamefont {P\"urrer}, \citenamefont {Jim\'nez~Forteza},\ and\
  \citenamefont {Boh\'e}}]{Khan:2015jqa}%
  \BibitemOpen
  \bibfield  {author} {\bibinfo {author} {\bibfnamefont {S.}~\bibnamefont
  {Khan}}, \bibinfo {author} {\bibfnamefont {S.}~\bibnamefont {Husa}}, \bibinfo
  {author} {\bibfnamefont {M.}~\bibnamefont {Hannam}}, \bibinfo {author}
  {\bibfnamefont {F.}~\bibnamefont {Ohme}}, \bibinfo {author} {\bibfnamefont
  {M.}~\bibnamefont {P\"urrer}}, \bibinfo {author} {\bibfnamefont
  {X.}~\bibnamefont {Jim\'nez~Forteza}},\ and\ \bibinfo {author} {\bibfnamefont
  {A.}~\bibnamefont {Boh\'e}},\ }\bibfield  {title} {\bibinfo {title}
  {{Frequency-domain gravitational waves from nonprecessing black-hole
  binaries. II. A phenomenological model for the advanced detector era}},\
  }\href {https://doi.org/10.1103/PhysRevD.93.044007} {\bibfield  {journal}
  {\bibinfo  {journal} {Phys. Rev.}\ }\textbf {\bibinfo {volume} {D93}},\
  \bibinfo {pages} {044007} (\bibinfo {year} {2016})},\ \Eprint
  {https://arxiv.org/abs/1508.07253} {arXiv:1508.07253 [gr-qc]} \BibitemShut
  {NoStop}%
\bibitem [{\citenamefont {Hannam}\ \emph {et~al.}(2014)\citenamefont {Hannam},
  \citenamefont {Schmidt}, \citenamefont {Bohé}, \citenamefont {Haegel},
  \citenamefont {Husa}, \citenamefont {Ohme}, \citenamefont {Pratten},\ and\
  \citenamefont {Pürrer}}]{Hannam:2013oca}%
  \BibitemOpen
  \bibfield  {author} {\bibinfo {author} {\bibfnamefont {M.}~\bibnamefont
  {Hannam}}, \bibinfo {author} {\bibfnamefont {P.}~\bibnamefont {Schmidt}},
  \bibinfo {author} {\bibfnamefont {A.}~\bibnamefont {Bohé}}, \bibinfo
  {author} {\bibfnamefont {L.}~\bibnamefont {Haegel}}, \bibinfo {author}
  {\bibfnamefont {S.}~\bibnamefont {Husa}}, \bibinfo {author} {\bibfnamefont
  {F.}~\bibnamefont {Ohme}}, \bibinfo {author} {\bibfnamefont {G.}~\bibnamefont
  {Pratten}},\ and\ \bibinfo {author} {\bibfnamefont {M.}~\bibnamefont
  {Pürrer}},\ }\bibfield  {title} {\bibinfo {title} {{Simple Model of Complete
  Precessing Black-Hole-Binary Gravitational Waveforms}},\ }\href
  {https://doi.org/10.1103/PhysRevLett.113.151101} {\bibfield  {journal}
  {\bibinfo  {journal} {Phys. Rev. Lett.}\ }\textbf {\bibinfo {volume} {113}},\
  \bibinfo {pages} {151101} (\bibinfo {year} {2014})},\ \Eprint
  {https://arxiv.org/abs/1308.3271} {arXiv:1308.3271 [gr-qc]} \BibitemShut
  {NoStop}%
\bibitem [{\citenamefont {Boh\'e}\ \emph {et~al.}(2017)\citenamefont {Boh\'e}
  \emph {et~al.}}]{Bohe:2016gbl}%
  \BibitemOpen
  \bibfield  {author} {\bibinfo {author} {\bibfnamefont {A.}~\bibnamefont
  {Boh\'e}} \emph {et~al.},\ }\bibfield  {title} {\bibinfo {title} {{Improved
  effective-one-body model of spinning, nonprecessing binary black holes for
  the era of gravitational-wave astrophysics with advanced detectors}},\ }\href
  {https://doi.org/10.1103/PhysRevD.95.044028} {\bibfield  {journal} {\bibinfo
  {journal} {Phys. Rev. D}\ }\textbf {\bibinfo {volume} {95}},\ \bibinfo
  {pages} {044028} (\bibinfo {year} {2017})},\ \Eprint
  {https://arxiv.org/abs/1611.03703} {arXiv:1611.03703 [gr-qc]} \BibitemShut
  {NoStop}%
\bibitem [{\citenamefont {Veitch}\ and\ \citenamefont
  {Del~Pozzo}(2013)}]{veitch2013analytic}%
  \BibitemOpen
  \bibfield  {author} {\bibinfo {author} {\bibfnamefont {J.}~\bibnamefont
  {Veitch}}\ and\ \bibinfo {author} {\bibfnamefont {W.}~\bibnamefont
  {Del~Pozzo}},\ }\bibfield  {title} {\bibinfo {title} {Analytic
  marginalisation of phase parameter},\ }\href@noop {} {\bibfield  {journal}
  {\bibinfo  {journal} {URL: https://dcc. ligo. org/LIGO-T1300326/public}\ }
  (\bibinfo {year} {2013})}\BibitemShut {NoStop}%
\bibitem [{\citenamefont {Thrane}\ and\ \citenamefont
  {Talbot}(2019)}]{Thrane:2018qnx}%
  \BibitemOpen
  \bibfield  {author} {\bibinfo {author} {\bibfnamefont {E.}~\bibnamefont
  {Thrane}}\ and\ \bibinfo {author} {\bibfnamefont {C.}~\bibnamefont
  {Talbot}},\ }\bibfield  {title} {\bibinfo {title} {{An introduction to
  Bayesian inference in gravitational-wave astronomy: parameter estimation,
  model selection, and hierarchical models}},\ }\href
  {https://doi.org/10.1017/pasa.2019.2} {\bibfield  {journal} {\bibinfo
  {journal} {Publ. Astron. Soc. Austral.}\ }\textbf {\bibinfo {volume} {36}},\
  \bibinfo {pages} {e010} (\bibinfo {year} {2019})},\ \bibinfo {note}
  {[Erratum: Publ.Astron.Soc.Austral. 37, e036 (2020)]},\ \Eprint
  {https://arxiv.org/abs/1809.02293} {arXiv:1809.02293 [astro-ph.IM]}
  \BibitemShut {NoStop}%
\bibitem [{\citenamefont {Lin}(1991)}]{Lin:1991zzm}%
  \BibitemOpen
  \bibfield  {author} {\bibinfo {author} {\bibfnamefont {J.}~\bibnamefont
  {Lin}},\ }\bibfield  {title} {\bibinfo {title} {{Divergence measures based on
  the Shannon entropy}},\ }\href {https://doi.org/10.1109/18.61115} {\bibfield
  {journal} {\bibinfo  {journal} {IEEE Trans. Info. Theor.}\ }\textbf {\bibinfo
  {volume} {37}},\ \bibinfo {pages} {145} (\bibinfo {year} {1991})}\BibitemShut
  {NoStop}%
\bibitem [{\citenamefont {{Romero-Shaw}}\ \emph {et~al.}(2020)\citenamefont
  {{Romero-Shaw}}, \citenamefont {{Talbot}}, \citenamefont {{Biscoveanu}},
  \citenamefont {{D'Emilio}}, \citenamefont {{Ashton}}, \citenamefont
  {{Berry}}, \citenamefont {{Coughlin}}, \citenamefont {{Galaudage}},
  \citenamefont {{Hoy}}, \citenamefont {{H{\"u}bner}}, \citenamefont
  {{Phukon}}, \citenamefont {{Pitkin}}, \citenamefont {{Rizzo}}, \citenamefont
  {{Sarin}}, \citenamefont {{Smith}}, \citenamefont {{Stevenson}},
  \citenamefont {{Vajpeyi}}, \citenamefont {{Ar{\`e}ne}}, \citenamefont
  {{Athar}}, \citenamefont {{Banagiri}}, \citenamefont {{Bose}}, \citenamefont
  {{Carney}}, \citenamefont {{Chatziioannou}}, \citenamefont {{Clark}},
  \citenamefont {{Colleoni}}, \citenamefont {{Cotesta}}, \citenamefont
  {{Edelman}}, \citenamefont {{Estell{\'e}s}}, \citenamefont
  {{Garc{\'\i}a-Quir{\'o}s}}, \citenamefont {{Ghosh}}, \citenamefont {{Green}},
  \citenamefont {{Haster}}, \citenamefont {{Husa}}, \citenamefont {{Keitel}},
  \citenamefont {{Kim}}, \citenamefont {{Hernandez-Vivanco}}, \citenamefont
  {{Maga{\~n}a Hernandez}}, \citenamefont {{Karathanasis}}, \citenamefont
  {{Lasky}}, \citenamefont {{De Lillo}}, \citenamefont {{Lower}}, \citenamefont
  {{Macleod}}, \citenamefont {{Mateu-Lucena}}, \citenamefont {{Miller}},
  \citenamefont {{Millhouse}}, \citenamefont {{Morisaki}}, \citenamefont
  {{Oh}}, \citenamefont {{Ossokine}}, \citenamefont {{Payne}}, \citenamefont
  {{Powell}}, \citenamefont {{Pratten}}, \citenamefont {{P{\"u}rrer}},
  \citenamefont {{Ramos-Buades}}, \citenamefont {{Raymond}}, \citenamefont
  {{Thrane}}, \citenamefont {{Veitch}}, \citenamefont {{Williams}},
  \citenamefont {{Williams}},\ and\ \citenamefont
  {{Xiao}}}]{2020MNRAS.499.3295R}%
  \BibitemOpen
  \bibfield  {author} {\bibinfo {author} {\bibfnamefont {I.~M.}\ \bibnamefont
  {{Romero-Shaw}}}, \bibinfo {author} {\bibfnamefont {C.}~\bibnamefont
  {{Talbot}}}, \bibinfo {author} {\bibfnamefont {S.}~\bibnamefont
  {{Biscoveanu}}}, \bibinfo {author} {\bibfnamefont {V.}~\bibnamefont
  {{D'Emilio}}}, \bibinfo {author} {\bibfnamefont {G.}~\bibnamefont
  {{Ashton}}}, \bibinfo {author} {\bibfnamefont {C.~P.~L.}\ \bibnamefont
  {{Berry}}}, \bibinfo {author} {\bibfnamefont {S.}~\bibnamefont {{Coughlin}}},
  \bibinfo {author} {\bibfnamefont {S.}~\bibnamefont {{Galaudage}}}, \bibinfo
  {author} {\bibfnamefont {C.}~\bibnamefont {{Hoy}}}, \bibinfo {author}
  {\bibfnamefont {M.}~\bibnamefont {{H{\"u}bner}}}, \bibinfo {author}
  {\bibfnamefont {K.~S.}\ \bibnamefont {{Phukon}}}, \bibinfo {author}
  {\bibfnamefont {M.}~\bibnamefont {{Pitkin}}}, \bibinfo {author}
  {\bibfnamefont {M.}~\bibnamefont {{Rizzo}}}, \bibinfo {author} {\bibfnamefont
  {N.}~\bibnamefont {{Sarin}}}, \bibinfo {author} {\bibfnamefont
  {R.}~\bibnamefont {{Smith}}}, \bibinfo {author} {\bibfnamefont
  {S.}~\bibnamefont {{Stevenson}}}, \bibinfo {author} {\bibfnamefont
  {A.}~\bibnamefont {{Vajpeyi}}}, \bibinfo {author} {\bibfnamefont
  {M.}~\bibnamefont {{Ar{\`e}ne}}}, \bibinfo {author} {\bibfnamefont
  {K.}~\bibnamefont {{Athar}}}, \bibinfo {author} {\bibfnamefont
  {S.}~\bibnamefont {{Banagiri}}}, \bibinfo {author} {\bibfnamefont
  {N.}~\bibnamefont {{Bose}}}, \bibinfo {author} {\bibfnamefont
  {M.}~\bibnamefont {{Carney}}}, \bibinfo {author} {\bibfnamefont
  {K.}~\bibnamefont {{Chatziioannou}}}, \bibinfo {author} {\bibfnamefont
  {J.~A.}\ \bibnamefont {{Clark}}}, \bibinfo {author} {\bibfnamefont
  {M.}~\bibnamefont {{Colleoni}}}, \bibinfo {author} {\bibfnamefont
  {R.}~\bibnamefont {{Cotesta}}}, \bibinfo {author} {\bibfnamefont
  {B.}~\bibnamefont {{Edelman}}}, \bibinfo {author} {\bibfnamefont
  {H.}~\bibnamefont {{Estell{\'e}s}}}, \bibinfo {author} {\bibfnamefont
  {C.}~\bibnamefont {{Garc{\'\i}a-Quir{\'o}s}}}, \bibinfo {author}
  {\bibfnamefont {A.}~\bibnamefont {{Ghosh}}}, \bibinfo {author} {\bibfnamefont
  {R.}~\bibnamefont {{Green}}}, \bibinfo {author} {\bibfnamefont {C.~J.}\
  \bibnamefont {{Haster}}}, \bibinfo {author} {\bibfnamefont {S.}~\bibnamefont
  {{Husa}}}, \bibinfo {author} {\bibfnamefont {D.}~\bibnamefont {{Keitel}}},
  \bibinfo {author} {\bibfnamefont {A.~X.}\ \bibnamefont {{Kim}}}, \bibinfo
  {author} {\bibfnamefont {F.}~\bibnamefont {{Hernandez-Vivanco}}}, \bibinfo
  {author} {\bibfnamefont {I.}~\bibnamefont {{Maga{\~n}a Hernandez}}}, \bibinfo
  {author} {\bibfnamefont {C.}~\bibnamefont {{Karathanasis}}}, \bibinfo
  {author} {\bibfnamefont {P.~D.}\ \bibnamefont {{Lasky}}}, \bibinfo {author}
  {\bibfnamefont {N.}~\bibnamefont {{De Lillo}}}, \bibinfo {author}
  {\bibfnamefont {M.~E.}\ \bibnamefont {{Lower}}}, \bibinfo {author}
  {\bibfnamefont {D.}~\bibnamefont {{Macleod}}}, \bibinfo {author}
  {\bibfnamefont {M.}~\bibnamefont {{Mateu-Lucena}}}, \bibinfo {author}
  {\bibfnamefont {A.}~\bibnamefont {{Miller}}}, \bibinfo {author}
  {\bibfnamefont {M.}~\bibnamefont {{Millhouse}}}, \bibinfo {author}
  {\bibfnamefont {S.}~\bibnamefont {{Morisaki}}}, \bibinfo {author}
  {\bibfnamefont {S.~H.}\ \bibnamefont {{Oh}}}, \bibinfo {author}
  {\bibfnamefont {S.}~\bibnamefont {{Ossokine}}}, \bibinfo {author}
  {\bibfnamefont {E.}~\bibnamefont {{Payne}}}, \bibinfo {author} {\bibfnamefont
  {J.}~\bibnamefont {{Powell}}}, \bibinfo {author} {\bibfnamefont
  {G.}~\bibnamefont {{Pratten}}}, \bibinfo {author} {\bibfnamefont
  {M.}~\bibnamefont {{P{\"u}rrer}}}, \bibinfo {author} {\bibfnamefont
  {A.}~\bibnamefont {{Ramos-Buades}}}, \bibinfo {author} {\bibfnamefont
  {V.}~\bibnamefont {{Raymond}}}, \bibinfo {author} {\bibfnamefont
  {E.}~\bibnamefont {{Thrane}}}, \bibinfo {author} {\bibfnamefont
  {J.}~\bibnamefont {{Veitch}}}, \bibinfo {author} {\bibfnamefont
  {D.}~\bibnamefont {{Williams}}}, \bibinfo {author} {\bibfnamefont {M.~J.}\
  \bibnamefont {{Williams}}},\ and\ \bibinfo {author} {\bibfnamefont
  {L.}~\bibnamefont {{Xiao}}},\ }\bibfield  {title} {\bibinfo {title}
  {{Bayesian inference for compact binary coalescences with BILBY: validation
  and application to the first LIGO-Virgo gravitational-wave transient
  catalogue}},\ }\href {https://doi.org/10.1093/mnras/staa2850} {\bibfield
  {journal} {\bibinfo  {journal} {MNRAS}\ }\textbf {\bibinfo {volume} {499}},\
  \bibinfo {pages} {3295} (\bibinfo {year} {2020})},\ \Eprint
  {https://arxiv.org/abs/2006.00714} {arXiv:2006.00714 [astro-ph.IM]}
  \BibitemShut {NoStop}%
\bibitem [{\citenamefont {Paszke}\ \emph {et~al.}(2019)\citenamefont {Paszke},
  \citenamefont {Gross}, \citenamefont {Massa}, \citenamefont {Lerer},
  \citenamefont {Bradbury}, \citenamefont {Chanan}, \citenamefont {Killeen},
  \citenamefont {Lin}, \citenamefont {Gimelshein}, \citenamefont {Antiga},
  \citenamefont {Desmaison}, \citenamefont {Kopf}, \citenamefont {Yang},
  \citenamefont {DeVito}, \citenamefont {Raison}, \citenamefont {Tejani},
  \citenamefont {Chilamkurthy}, \citenamefont {Steiner}, \citenamefont {Fang},
  \citenamefont {Bai},\ and\ \citenamefont {Chintala}}]{NEURIPS2019_9015}%
  \BibitemOpen
  \bibfield  {author} {\bibinfo {author} {\bibfnamefont {A.}~\bibnamefont
  {Paszke}}, \bibinfo {author} {\bibfnamefont {S.}~\bibnamefont {Gross}},
  \bibinfo {author} {\bibfnamefont {F.}~\bibnamefont {Massa}}, \bibinfo
  {author} {\bibfnamefont {A.}~\bibnamefont {Lerer}}, \bibinfo {author}
  {\bibfnamefont {J.}~\bibnamefont {Bradbury}}, \bibinfo {author}
  {\bibfnamefont {G.}~\bibnamefont {Chanan}}, \bibinfo {author} {\bibfnamefont
  {T.}~\bibnamefont {Killeen}}, \bibinfo {author} {\bibfnamefont
  {Z.}~\bibnamefont {Lin}}, \bibinfo {author} {\bibfnamefont {N.}~\bibnamefont
  {Gimelshein}}, \bibinfo {author} {\bibfnamefont {L.}~\bibnamefont {Antiga}},
  \bibinfo {author} {\bibfnamefont {A.}~\bibnamefont {Desmaison}}, \bibinfo
  {author} {\bibfnamefont {A.}~\bibnamefont {Kopf}}, \bibinfo {author}
  {\bibfnamefont {E.}~\bibnamefont {Yang}}, \bibinfo {author} {\bibfnamefont
  {Z.}~\bibnamefont {DeVito}}, \bibinfo {author} {\bibfnamefont
  {M.}~\bibnamefont {Raison}}, \bibinfo {author} {\bibfnamefont
  {A.}~\bibnamefont {Tejani}}, \bibinfo {author} {\bibfnamefont
  {S.}~\bibnamefont {Chilamkurthy}}, \bibinfo {author} {\bibfnamefont
  {B.}~\bibnamefont {Steiner}}, \bibinfo {author} {\bibfnamefont
  {L.}~\bibnamefont {Fang}}, \bibinfo {author} {\bibfnamefont {J.}~\bibnamefont
  {Bai}},\ and\ \bibinfo {author} {\bibfnamefont {S.}~\bibnamefont
  {Chintala}},\ }\bibfield  {title} {\bibinfo {title} {Pytorch: An imperative
  style, high-performance deep learning library},\ }in\ \href
  {http://papers.neurips.cc/paper/9015-pytorch-an-imperative-style-high-performance-deep-learning-library.pdf}
  {\emph {\bibinfo {booktitle} {Advances in Neural Information Processing
  Systems 32}}},\ \bibinfo {editor} {edited by\ \bibinfo {editor}
  {\bibfnamefont {H.}~\bibnamefont {Wallach}}, \bibinfo {editor} {\bibfnamefont
  {H.}~\bibnamefont {Larochelle}}, \bibinfo {editor} {\bibfnamefont
  {A.}~\bibnamefont {Beygelzimer}}, \bibinfo {editor} {\bibfnamefont
  {F.}~\bibnamefont {d'Alch\'{e} Buc}}, \bibinfo {editor} {\bibfnamefont
  {E.}~\bibnamefont {Fox}},\ and\ \bibinfo {editor} {\bibfnamefont
  {R.}~\bibnamefont {Garnett}}}\ (\bibinfo  {publisher} {Curran Associates,
  Inc.},\ \bibinfo {year} {2019})\ pp.\ \bibinfo {pages}
  {8024--8035}\BibitemShut {NoStop}%
\bibitem [{\citenamefont {Durkan}\ \emph {et~al.}(2020)\citenamefont {Durkan},
  \citenamefont {Bekasov}, \citenamefont {Murray},\ and\ \citenamefont
  {Papamakarios}}]{nflows}%
  \BibitemOpen
  \bibfield  {author} {\bibinfo {author} {\bibfnamefont {C.}~\bibnamefont
  {Durkan}}, \bibinfo {author} {\bibfnamefont {A.}~\bibnamefont {Bekasov}},
  \bibinfo {author} {\bibfnamefont {I.}~\bibnamefont {Murray}},\ and\ \bibinfo
  {author} {\bibfnamefont {G.}~\bibnamefont {Papamakarios}},\ }\href
  {https://doi.org/10.5281/zenodo.4296287} {\bibinfo {title} {{nflows}:
  normalizing flows in {PyTorch}}} (\bibinfo {year} {2020})\BibitemShut
  {NoStop}%
\bibitem [{\citenamefont {{LIGO Scientific Collaboration}}(2018)}]{lalsuite}%
  \BibitemOpen
  \bibfield  {author} {\bibinfo {author} {\bibnamefont {{LIGO Scientific
  Collaboration}}},\ }\href {https://doi.org/10.7935/GT1W-FZ16} {\bibinfo
  {title} {{LIGO} {A}lgorithm {L}ibrary - {LALS}uite}},\ \bibinfo
  {howpublished} {free software (GPL)} (\bibinfo {year} {2018})\BibitemShut
  {NoStop}%
\bibitem [{\citenamefont {Kingma}\ and\ \citenamefont
  {Ba}(2014)}]{Kingma:2014vow}%
  \BibitemOpen
  \bibfield  {author} {\bibinfo {author} {\bibfnamefont {D.~P.}\ \bibnamefont
  {Kingma}}\ and\ \bibinfo {author} {\bibfnamefont {J.}~\bibnamefont {Ba}},\
  }\bibfield  {title} {\bibinfo {title} {{Adam: A Method for Stochastic
  Optimization}},\ }\href@noop {} {\  (\bibinfo {year} {2014})},\ \Eprint
  {https://arxiv.org/abs/1412.6980} {arXiv:1412.6980 [cs.LG]} \BibitemShut
  {NoStop}%
\bibitem [{\citenamefont {Hunter}(2007)}]{Hunter:2007}%
  \BibitemOpen
  \bibfield  {author} {\bibinfo {author} {\bibfnamefont {J.~D.}\ \bibnamefont
  {Hunter}},\ }\bibfield  {title} {\bibinfo {title} {Matplotlib: A 2d graphics
  environment},\ }\href {https://doi.org/10.1109/MCSE.2007.55} {\bibfield
  {journal} {\bibinfo  {journal} {Computing in Science \& Engineering}\
  }\textbf {\bibinfo {volume} {9}},\ \bibinfo {pages} {90} (\bibinfo {year}
  {2007})}\BibitemShut {NoStop}%
\bibitem [{\citenamefont {{Hinton}}(2016)}]{Hinton2016}%
  \BibitemOpen
  \bibfield  {author} {\bibinfo {author} {\bibfnamefont {S.~R.}\ \bibnamefont
  {{Hinton}}},\ }\bibfield  {title} {\bibinfo {title} {{ChainConsumer}},\
  }\href {https://doi.org/10.21105/joss.00045} {\bibfield  {journal} {\bibinfo
  {journal} {The Journal of Open Source Software}\ }\textbf {\bibinfo {volume}
  {1}},\ \bibinfo {eid} {00045} (\bibinfo {year} {2016})}\BibitemShut {NoStop}%
\end{thebibliography}%

\end{document}